\documentclass[aps,pra,twocolumn,showpacs,unsortedaddress]{revtex4-1}

\usepackage{amsmath,amssymb}
\usepackage{graphicx}

\newcommand{\kc}{k}
\newcommand{\kac}{\kappa}

\newcommand{\rme}{\mathrm e}
\newcommand{\rmi}{\mathrm i}
\newcommand{\rmd}{{\mathrm d}}

\renewcommand{\Re}{\mathop{{\rm Re}}}
\renewcommand{\Im}{\mathop{{\rm Im}}}

\begin{document}

\title{Equilibration and thermalization of the dissipative quantum harmonic oscillator \\ in a non-thermal environment}

\author{D. Pagel}
\author{A. Alvermann}
\email{alvermann@physik.uni-greifswald.de}
\author{H. Fehske}
\affiliation{Institut f\"ur Physik, Ernst-Moritz-Arndt-Universit\"at, 17487 Greifswald, Germany}

\begin{abstract}
 We study the dissipative quantum harmonic oscillator with general non-thermal preparations of the harmonic oscillator bath. 
 The focus is on equilibration of the oscillator in the long-time limit and the additional requirements for thermalization.
 Our study is based on the exact solution of the microscopic model obtained by means of operator equations of motion,
which provides us with the time evolution of the central oscillator density matrix in terms of the propagating function. 
We find a hierarchy of conditions for thermalization,
together with the relation of the asymptotic temperature to the energy distribution in the initial bath state.
We discuss the presence and absence of equilibration for the example of an inhomogeneous chain of harmonic oscillators, and illustrate the general findings about thermalization for the non-thermal environment that results from a quench.

\end{abstract}


\maketitle

\section{Introduction}

Equilibration can be defined as the evolution of a 
system out of equilibrium towards a stationary state in the long-time limit.
For quantum systems, the question arises how equilibration is possible in spite of the linear and unitary time evolution,
how the stationary state depends on the initial conditions,
and to which extent it can be described as a thermal state.

General arguments relate equilibration to dephasing of quantum states~\cite{BS08,LPSW09,Rei08,Rei10,Yuk11}.
Starting from the expansion of an initial state $|\psi(0)\rangle  = \sum_{n=1}^N \psi_n |n\rangle$
in the eigenstates $|n\rangle$ of the Hamiltonian $H = \sum_{n=1}^N  E_n |n\rangle \langle n|$, 
the time evolution of an operator expectation value is given by 
\begin{eqnarray}
 \langle A(t) \rangle &=& \langle \psi(t) | A | \psi(t) \rangle \nonumber\\
 &=&  \sum_{m,n=1}^N \psi_m^* \psi_n \, \rme^{\rmi(E_m-E_n)t} \langle m | A |n \rangle \;.
\end{eqnarray}
In the thermodynamic limit \mbox{$N \to \infty$}
we can expect that only diagonal terms $m=n$ survive for  $t \to \infty$,
such that the long-time limit of the expectation value is 
\begin{equation}\label{GeneralA}
\lim_{t \to \pm \infty} \langle A(t) \rangle \simeq  \mathrm{tr} \big[  \rho_\infty A \big]  \quad (N \to \infty) \;,
\end{equation}
with the density matrix $\rho_\infty = \sum_{n=1}^N |\psi_n|^2 |n\rangle \langle n|$.
This argument can be justified with the Riemann-Lebesgue lemma~\cite{Koe89} that states
\begin{equation}\label{ftprop}
 \lim_{t \to \pm \infty} \int_{-\infty}^\infty f(\omega) \, \rme^{\rmi \omega t} \, \rmd \omega = 0 
\end{equation}
for any integrable function $f(\omega)$ (here: the density of states $D(\omega)$).
Although this argument explains the origin of equilibration,
not much is learned about the properties of the stationary state $\rho_\infty$.
Especially the question of thermalization is left open.

In this paper we study equilibration and thermalization of dissipative quantum harmonic oscillators, using the standard model of a central oscillator coupled to a harmonic oscillator bath. For this example we can determine the stationary state $\rho_\infty$ explicitly and analyze its dependence on the initial conditions completely.
Crucially, we allow for arbitrary non-thermal bath preparations in our study.
Thermalization is subject to additional conditions in this more general situation, and we show how the temperature of the asymptotic stationary state is obtained from the initial energy distribution of the oscillator bath rather than from the initial bath temperature.
We also include the case study of an interaction quench in an infinite harmonic chain, where undamped oscillations can prevent equilibration at strong damping.

The dissipative quantum harmonic oscillator is studied extensively in the literature~\cite{HR85,FLC85,FLC88,FC07},
covering such diverse topics as Brownian motion~\cite{Ull66a,Ull66b,Ull66c,Ull66d,Gra06}, quantum fluctuations~\cite{NG02}, driven dissipative systems~\cite{ZH95},
entanglement~\cite{JB04}, the existence of local temperatures~\cite{HMH04}, or the second law of thermodynamics~\cite{KM10}.
Reviews are given, e.g., in~\cite{GSI88,Wei99,FRH11}.
With an exact solution this model is also an important example for the derivation of master equations~\cite{UZ89,HPZ92,KG97,BP02}, the discussion of fundamental statistical relations such as
fluctuation-dissipation theorems~\cite{CHT11}
 and their connection to detailed balance and the Kubo-Martin-Schwinger condition~\cite{FH12},
 or for the assessment of numerical methods that provide a perspective for non-linear models~\cite{TRJF98,TRH00}.
It appears, however, that the questions addressed here have not been previously analyzed in detail, especially not for non-thermal bath preparations.

To obtain our results we proceed as follows.
After introduction of the model in Sec.~\ref{sec:model},
we construct the exact solution for non-thermal initial states in Sec.~\ref{sec:EOM},
including the propagating function in Sec.~\ref{sec:prop}.
Further details, including the extension to driven oscillators, are given in App.~\ref{app:Deriv} and App.~\ref{app:Prop}.
The central results for equilibration and thermalization are formulated in Sec.~\ref{sec:Equil}.
We discuss these results for the example of an infinite chain of harmonic oscillators in Sec.~\ref{sec:Chain},
before we conclude in Sec.~\ref{sec:Summary}.

\section{\label{sec:model}The model}

The Hamiltonian for the dissipative quantum harmonic oscillator, 
\begin{equation}\label{Ham}
 H = H_S + H_B + H_{SB} \;,
\end{equation}
is the sum of the contribution of the central oscillator,
\begin{equation}\label{HamS}
 H_S = \frac 1 2 \Big[ P^2 + \Omega^2 Q^2 \Big] \;,
 \end{equation}
 the contribution of the harmonic oscillator bath, 
 \begin{equation}\label{HB}
 H_B = \frac 1 2 \sum_{\nu=1}^N \Big[ P^2_\nu + \omega_\nu^2 Q_\nu^2 \Big] \;,
 \end{equation}
 and the linear interaction term
 \begin{equation} \label{HSB}
 H_{SB} = Q \sum_{\nu=1}^N \lambda_\nu Q_\nu \;.
\end{equation}
In these expressions, $Q_\nu$, $P_\nu$ are position and momentum operators
with canonical commutation relations, e.g. $[Q_\mu,P_\nu]=\rmi \delta_{\mu \nu}$.
Summations over Greek indices, used for bath oscillator operators $Q_\nu$, $P_\nu$,
run from $1, \dots, N$.
We suppress an index  for the central oscillator operators.

The size of the coupling constants $\lambda_\nu$ is restricted by the 
positivity condition
\begin{equation}\label{pos}
 \Omega^2 - \sum_{\nu=1}^N \frac{\lambda_\nu^2}{\omega_\nu^2} \geq 0 \;.
\end{equation}
It guarantees that the normal modes of the total Hamiltonian $H$ have real frequencies,
such that $H$ is bounded from below~\cite{Ull66a}.
A positive Hamiltonian can always be obtained through addition of the term 
$(1/2) \sum_{\nu=1}^N (\lambda_\nu/\omega_\nu)^2 Q^2$,
which leads to renormalization of the central oscillator frequency~\cite{Wei99}.
We prefer the present form of the Hamiltonian since it allows for a more natural treatment of the harmonic chain in Sec.~\ref{sec:Chain}.

Of primary interest to us is the central oscillator density matrix
\begin{equation}
\rho_S(t) = \mathrm{tr}_B [ \exp(-\rmi H t)  \rho(0) \exp( \rmi H t) ] \;,
\end{equation}
which is obtained from the initial state $\rho(0)$ through propagation with the total Hamiltonian $H$ and subsequent evaluation of the partial trace $\mathrm{tr}_B$ over the bath degrees of freedom.
A natural choice for $\rho(0)$ are factorizing initial conditions
\begin{equation}\label{RhoInitial}
 \rho(0) = \rho_S(0) \otimes \rho_B(0) \;,
\end{equation}
which correspond to the picture that at $t=0$ the previously isolated central oscillator is brought into contact with the oscillator bath.

The restriction to factorizing initial conditions is not essential for the following derivations,
especially not for the long-time limit in Sec.~\ref{sec:Equil},
but it is a natural assumption that simplifies the presentation. For example, mixed central/bath oscillator terms drop out of the expressions for the central oscillator variance (see Sec.~\ref{sec:ExpValues}).

\section{\label{sec:EOM}Solution of the dissipative quantum oscillator for general initial conditions}

The central oscillator density matrix $\rho_S(t)$ can be obtained in various ways,
e.g. through transformation of $H$ to normal modes~\cite{Ull66a,HR85}
or by using path integrals~\cite{GWT84,GSI88} based on the Feynman-Vernon influence functional formalism~\cite{FV63,CL83a,CL83b}.
The arguably simplest approach is the direct solution of the Heisenberg equations of motion for 
the operators $Q(t)$, $P(t)$, which reduces to the solution of a classical equation of motion.
The initial conditions $\rho_S(0)$ and $\rho_B(0)$ enter only the evaluation of central oscillator expectation values, such that we can allow for general initial bath states.
The full solution is then given by the propagating function.

\subsection{Reduction to classical equation of motion}

As further detailed in App.~\ref{app:Deriv}, the central piece of information is the solution 
$u(t) \in \mathbb{R}$ of the classical equation of motion
\begin{equation}\label{uhom}
 \ddot{u}(t) = - \Omega^2 u(t) + \int_{0}^{t} K(t-\tau) u(\tau) \, \rmd\tau \;,
\end{equation}
which is subject to the conditions
\begin{enumerate}
\item
 $u(t)$ solves Eq.~\eqref{uhom} for $t>0$,
 \item
 $u(t)=0$ for $t<0$,
\item
the initial conditions are $u(0)=0$, $\dot{u}(0)=1$.
\end{enumerate}
We here introduced the damping kernel
\begin{equation}\label{DampKern}
 K(t) = \sum\limits_{\nu=1}^N \frac{\lambda_\nu^2}{\omega_\nu} \sin \omega_\nu t \;.
\end{equation}
The function $u(t)$ can be calculated as the Fourier transform~\cite{Dav02}
\begin{equation}\label{UFromFourier2}
 u(t) = \frac{2}{\pi} \int_0^\infty \sin \omega t \Im F(\omega + \rmi 0^+) \, \rmd\omega 
\end{equation}
of the function 
\begin{equation}
 F(z) = \Big(  \Omega^2  - z^2  + \sum_{\nu=1}^N \frac{\lambda_\nu^2}{z^2 - \omega_\nu^2} \Big)^{-1} \; ,
\end{equation}
writing $F(\omega + \rmi 0^+) = \lim_{\eta \to 0, \eta>0} F(\omega + \rmi \eta)$.
We note that the positivity condition~\eqref{pos} implies that the poles of $F(z)$ occur on the real axis, such that $u(t)$ is a quasiperiodic function for finite $N$ while $u(t) \to 0$ for $t \to \infty$ is possible in the thermodynamic limit $N \to \infty$.
An explicit example for the computation of $u(t)$ is given for the harmonic chain in Sec.~\ref{sec:Chain} (see Eq.~\eqref{UChain}).

To proceed, we introduce the partial Fourier transforms
\begin{equation}\label{UFour}
 \tilde{u}(t, \omega) = \rme^{\rmi t \omega} \int_0^t u(\tau) \, \rme^{- \rmi  \omega \tau} \, \rmd\tau \;,
\end{equation}
\begin{equation}\label{UFourDot}
 \tilde{v}(t, \omega) = \rme^{\rmi t \omega} \int_0^t \dot{u} (\tau) \, \rme^{- \rmi \omega \tau} \, \rmd\tau
 = u(t) + \rmi \omega \tilde{u}(t, \omega) \;,
\end{equation}
and define the matrices
\begin{equation}\label{UMat}
 {\mathbf U}(t) = \begin{pmatrix} U_{QQ}(t) & U_{QP}(t) \\ U_{PQ}(t) & U_{PP}(t) \end{pmatrix} = \begin{pmatrix} \dot{u}(t) & u(t) \\ \ddot{u}(t) & \dot{u}(t) \end{pmatrix} \;,
\end{equation}
\begin{equation}\label{UMatOm}
 {\mathbf U}(t, \omega) =   \begin{pmatrix} \Re \tilde{u}(t, \omega) & \dfrac{\Im \tilde{u}(t, \omega)}{\omega} \\[2ex]  \Re \tilde{v}(t, \omega) & \dfrac{\Im \tilde{v}(t, \omega)}{\omega}  \end{pmatrix} \;.
\end{equation}
We now obtain the central oscillator operators from the matrix equation
\begin{equation}\label{QSolMat}
 \begin{pmatrix} Q(t) \\ P(t) \end{pmatrix}
 = \mathbf{U}(t) \begin{pmatrix} Q(0) \\ P(0) \end{pmatrix}
 - \sum_{\nu=1}^N  \lambda_\nu \mathbf{U}(t,\omega_\nu)  \begin{pmatrix} Q_\nu(0) \\ P_\nu(0) \end{pmatrix} \;.
\end{equation}

\subsection{\label{sec:ExpValues}Central oscillator expectation values}

Eq.~\eqref{QSolMat} gives the operators $Q(t)$, $P(t)$ as linear combinations of the operators $Q(0), P(0)$ and $Q_\nu(0), P_\nu(0)$.
This allows us to express central oscillator expectation values for $t\ge 0$
in terms of the initial expectation values at $t=0$. 

The linear expectation values are given by the matrix equation 
\begin{equation}\label{X}
 {\mathbf X}(t) \equiv  \begin{pmatrix} \langle Q(t) \rangle \\[0.5ex] \langle P(t) \rangle \end{pmatrix} = {\mathbf U}(t) {\mathbf X}(0) + {\mathbf I}(t) \,,
\end{equation}
with the same shape as Eq.~\eqref{QSolMat}.
In addition to the initial expectation values $\mathbf{X}(0)$ it contains the contribution
\begin{equation}\label{I}
 {\mathbf I}(t) =  \begin{pmatrix} I_Q(t) \\ I_P(t) \end{pmatrix} = - \sum_{\nu=1}^N \lambda_\nu {\mathbf U}(t, \omega_\nu) \breve{\mathbf X}_\nu \,,
\end{equation}
 where we mark the initial bath expectation values
\begin{equation}\label{XBathInitial}
\breve{\mathbf X}_\nu = \begin{pmatrix} \langle Q_\nu(0) \rangle \\  \langle P_\nu(0) \rangle \end{pmatrix}
\end{equation} 
with a breve $\breve{\phantom{x}}$ as a notational convention.
Note that if $\breve{{\mathbf X}}_\nu \equiv 0$, e.g. for a thermal bath, the `noise term' $\mathbf{I}(t)$ vanishes. Then, position $\langle Q(t) \rangle$ and momentum $\langle P(t) \rangle$ of the central oscillator follow the classical equation of motion~\eqref{uhom}.

For the quadratic expectation values we define the variance of operators $A$, $B$ as
\begin{equation}
 \Sigma_{AB} = \frac{1}{2} \langle A B + B A \rangle - \langle A \rangle \langle B \rangle \;,
\end{equation}
which simplifies to $\Sigma_{AA} = \langle A^2 \rangle -\langle A \rangle^2$ for $A=B$,
and write $\Sigma_{AB}(t) = \Sigma_{A(t) B(t)}$.
We combine the central oscillator variances into the real symmetric matrix 
\begin{equation}\label{SigmaMatrix}
 \mathbf{\Sigma}(t) = \begin{pmatrix} \Sigma_{Q Q}(t) & \Sigma_{Q P}(t) \\[1ex] \Sigma_{Q P}(t) & \Sigma_{P P}(t) \end{pmatrix} \;,
\end{equation}
and denote the initial bath variances with the matrix
\begin{equation}\label{Sigma_numu}
 \breve{\mathbf{\Sigma}}_{\nu\mu} = \begin{pmatrix} \Sigma_{Q_\nu Q_\mu}(0) & \Sigma_{Q_\nu P_\mu}(0) \\[1ex] \Sigma_{Q_\mu P_\nu}(0)  & \Sigma_{P_\nu P_\mu}(0) 
 \end{pmatrix} \;.
\end{equation}
Note the index swap in the off-diagonal elements,
and recall that mixed central oscillator/bath variances such as $\Sigma_{Q Q_\nu}$ vanish for our choice~\eqref{RhoInitial} of factorizing initial conditions.

We now obtain with Eq.~\eqref{QSolMat} the matrix equation
\begin{equation}\label{Sigma}
 \mathbf{\Sigma}(t) = {\mathbf U}(t) \mathbf{\Sigma}(0) {\mathbf U}^T(t)  +  {\mathbf C}(t) \;. 
\end{equation}
Similar to Eq.~\eqref{X}, the first term results from the time evolution of the central oscillator according to the classical equation of motion~\eqref{uhom},
and appears in the same form for an isolated oscillator.
Only the second term 
\begin{eqnarray}\label{C}
 {\mathbf C}(t) &=& \begin{pmatrix} C_{QQ}(t) & C_{QP}(t) \\ C_{QP}(t) & C_{PP}(t) \end{pmatrix} \nonumber\\
 &=& \sum_{\nu, \mu=1}^N \lambda_\nu \lambda_\mu {\mathbf U}(t, \omega_\nu) \breve{\mathbf{\Sigma}}_{\nu\mu} {\mathbf U}^T(t, \omega_\mu) 
\end{eqnarray}
depends on the initial bath oscillator variances $\breve{\mathbf{\Sigma}}_{\nu\mu}$.
Mixed terms in $\mathbf{U}(t)$, $\mathbf{U}(t,\omega_\nu)$ do not appear for factorizing initial conditions.

\subsection{\label{sec:TDLimit}The thermodynamic limit}

Because $u(t)$ is a quasi-periodic function for a finite number $N$ of bath oscillators,
equilibration becomes possible only in the thermodynamic limit $N \to \infty$.
We assume that for $N \to \infty$ the density of states
\begin{equation}\label{D}
 D(\omega) = \frac{1}{N} \sum_{\nu=1}^N \delta(\omega - \omega_\nu)
\end{equation}
converges to a continuous function.
Note that $D(\omega)=0$ for $\omega<0$ since the bath oscillator frequencies are positive.
The coupling constants 
appear in the damping kernel $K(t)$ and in Eq.~\eqref{uhom}
as $\lambda_\nu^2$, and must thus scale as $N^{-1/2}$.
We assume that 
\begin{equation} \label{LambdaCont}
\lambda_\nu = \lambda(\omega_\nu) / \sqrt{N}
\end{equation} 
with a continuous function $\lambda(\omega)$, and introduce the bath spectral function
\begin{equation}\label{GammaCont}
 \gamma(\omega) = D(\omega) \frac{\lambda(\omega)^2}{\omega} \,,
\end{equation}
with $\gamma(\omega)=0$ for $\omega < 0$.
The damping kernel is now given as
\begin{equation}\label{Kcont}
 K(t) = \int_0^\infty \gamma(\omega) \sin \omega t \, \rmd\omega \,,
\end{equation}
and the positivity condition reads
\begin{equation}\label{PosCont}
 \Omega^2 \ge \int_0^\infty \frac{\gamma(\omega)}{\omega} \, \rmd\omega \;.
\end{equation}

The function $F(z)$ in Eq.~\eqref{UFromFourier2} for $u(t)$ can be written as
\begin{equation}\label{FCont}
 F(z) = \Big(  \Omega^2  - z^2  + \int_0^\infty \frac{\omega \gamma(\omega)}{z^2 - \omega^2} \, \rmd\omega \Big)^{-1} \;.
\end{equation}
Under mild assumptions, the evaluation of the $\omega$-integral in this equation is possible by contour integration and results in
\begin{equation}\label{FAnalytic}
  F(z) = \Big(  \Omega^2  - z^2  + \Gamma(z) \Big)^{-1}
\end{equation}
for $\Im z > 0$,
where the complex function $\Gamma(z)$ with
$\gamma( \omega) = \mp (2/\pi) \Im \Gamma( \pm \omega + \rmi 0^+)$ is the analytic continuation of $\gamma(\omega)$ into the upper half of the complex plane (see Sec.~\ref{sec:Chain} for an example). For future use in Sec.~\ref{sec:Equil} we note the relation  
$\gamma(\omega) |F(\omega)|^2 = (2/\pi) \Im F(\omega + \rmi 0^+)$ that follows from this representation.
The analytic properties of $F(z)$ determine the behavior of $u(t)$ in the long-time limit, which is essential for equilibration (see condition (E0) in Sec.~\ref{sec:Equil}): It is $u(t) \to 0$ for $t \to \infty$ if and only if $F(z)$ has no isolated poles.

The linear expectation values $\breve{\mathbf{X}}_\nu$ enter Eq.~\eqref{I} with the prefactors $\lambda_\nu \propto N^{-1/2}$.
To obtain a finite result for the sum over $N$ terms,
also $\breve{\mathbf{X}}_\nu$ has to scale as $N^{-1/2}$, which leads to the ansatz
\begin{equation}
 \breve{{\mathbf X}}_\nu = \frac{1}{\sqrt{N}} \breve{{\mathbf X}}(\omega_\nu) 
\end{equation}
with a continuous vector-valued function $\breve{\mathbf{X}}(\omega)$.
Then, Eq.~\eqref{I} becomes
\begin{equation}\label{ICont}
 {\mathbf I}(t) = - \int_0^\infty D(\omega) \lambda(\omega) {\mathbf U}(t, \omega)\breve{{\mathbf X}}(\omega) \, \rmd\omega \;.
\end{equation}

The variances $\breve{\mathbf{\Sigma}}_{\nu \mu}$ enter the sum in Eq.~\eqref{C} with the prefactors $\lambda_\nu \lambda_\mu \propto N^{-1}$.
We must now distinguish between the $N^2$ off-diagonal terms $\nu \ne \mu$,
which require an additional $1/N$ prefactor for convergence,
and the $N$ diagonal terms $\nu = \mu$.
Therefore, we make the ansatz
\begin{equation}\label{SigmaAnsatz}
 \breve{\mathbf{\Sigma}}_{\nu \mu} = \frac{1}{N} \breve{\mathbf{\Sigma}}^{(2)}(\omega_\nu,\omega_\mu) 
 + \breve{\mathbf{\Sigma}}^{(1)}(\omega_\nu) \delta_{\nu \mu}
\end{equation}
with continuous matrix-valued functions $\breve{\mathbf{\Sigma}}^{(2)}(\omega_1,\omega_2)$ and $\breve{\mathbf{\Sigma}}^{(1)}(\omega)$.
Then, $\mathbf{C}(t)$ from Eq.~\eqref{C} is the sum of 
the off-diagonal term
\begin{eqnarray}\label{CContOD}
   {\mathbf C}^{(2)}(t) &=& \iint_{0}^{\infty} D(\omega_1) D(\omega_2) \lambda(\omega_1) \lambda(\omega_2) \nonumber\\
  &&\times {\mathbf U}(t, \omega_1) \breve{\mathbf{\Sigma}}^{(2)}(\omega_1,\omega_2) {\mathbf U}^T(t, \omega_2) \, \rmd\omega_1 \, \rmd\omega_2 \qquad
\end{eqnarray}
and the diagonal term
\begin{equation}\label{CCont}
 {\mathbf C}^{(1)}(t) = \int_0^\infty \omega \gamma(\omega) {\mathbf U}(t, \omega) \breve{\mathbf{\Sigma}}^{(1)}(\omega) {\mathbf U}^T(t, \omega) \, \rmd\omega \,.
\end{equation}
If the initial bath state is uncorrelated, 
such as for a thermal bath or a general product state $\rho_B(0) = \rho_B^1(0) \otimes \cdots \otimes \rho_B^N(0)$,
the off-diagonal term $\mathbf{C}^{(2)}(t)$ vanishes.

When we construct the propagating function in the next subsection,
we will conveniently assume that the initial bath state $\rho_B(0)$ is a Gaussian state.
For the long-time limit, the situation of interest here, 
this assumption can be justified in the thermodynamic limit on general grounds~\cite{CDEO08,CE10}.
The principal mechanism is illustrated with counting arguments of the following kind:
Consider an uncorrelated bath state, where only $N$ diagonal terms contribute in any sum over the bath oscillators. If we consider a higher order cumulant of bath operators, say $Q_3(\nu) = \langle Q_\nu^3 \rangle - 3 \langle Q_\nu^2\rangle \langle Q_\nu \rangle + 2 \langle Q_\nu \rangle^3$ as mentioned before, it appears with a prefactor $\lambda_\nu^3 \propto N^{-3/2}$.
Therefore, the total contribution of these cumulants scales as $N \times N^{-3/2} = N^{-1/2}$
and vanishes in the limit $N \to \infty$. 
Similar counting arguments can be given for cumulants involving two or more bath oscillators in the presence of correlations.
Because higher order cumulants vanish and only linear and quadratic bath expectation values survive 
the $N  \to \infty$ and $t \to \infty$ limit, we can treat the bath state as Gaussian in any calculation of the central oscillator density matrix.
For the formulation and proof of a strict result, which is involved even under some simplifying assumptions, see~\cite{CE10}.

\subsection{\label{sec:prop}The propagating function}

Knowledge of the expectation values $\mathbf X(t)$, $\mathbf \Sigma(t)$ does not suffice to obtain the central oscillator density matrix $\rho_S(t)$, unless we 
restrict ourselves completely to Gaussian oscillator states (cf. Eq.~\eqref{GaussianWigner} below).
Otherwise, the general solution is given by the propagating function $J(\cdot)$ that, in position representation, expresses the density matrix $\rho_S(q,q',t) = \langle q|\rho_S(t) |q' \rangle$ for $t \ge 0$ as 
\begin{equation}\label{JDef}
 \rho_S(q_f, q_f', t) = \iint_{-\infty}^\infty \! J(q_f, q_f', q_i, q_i', t) \rho_S(q_i, q_i', 0) \, \rmd q_i \, \rmd q_i'  \;.
\end{equation}
This expression must hold for all $\rho_S(0)$ and $t \ge 0$, and a fixed initial bath state $\rho_B(0)$.

The propagating function can be calculated using path integrals 
and the result for a thermal bath is given, e.g., in~\cite{GSI88}. 
Within our approach it is more natural to construct the propagating function 
directly, using only that an initial Gaussian state of the joint central/bath oscillator system remains a Gaussian state during time evolution with the bilinear Hamiltonian $H$.
With respect to the final remarks in Sec.~\ref{sec:TDLimit}, we assume a Gaussian bath state $\rho_B(0)$.
We can then consider the most general ansatz for $J(\cdot)$ that maps an initial Gaussian state $\rho_S(0)$ in Eq.~\eqref{JDef} 
onto a Gaussian state $\rho_S(t)$ for $t \ge 0$,
and will find that the parameters of this ansatz are fully specified through the linear maps~\eqref{X},~\eqref{Sigma} of $\mathbf X(t)$, $\mathbf \Sigma(t)$.
The result is valid for arbitrary $\rho_S(0)$ in Eq.~\eqref{JDef},
but we do not need to consider non-Gaussian $\rho_S(t)$ explicitly.

To translate this argument into equations we work with the Wigner function~\cite{Wig32,Schl01}
\begin{equation}
 W(q,p,t) = \frac{1}{2\pi} \int_{-\infty}^\infty \rho_S \Big(q+\frac{s}{2},q-\frac{s}{2},t \Big) \, \rme^{-\rmi p s} \, \rmd s 
\end{equation}
instead of the density matrix $\rho_S(q,q',t)$ in position representation
(see also Refs.~\cite{CRV03,FRH11} for a related calculation).
The propagating function $J_W(\tilde{\mathbf{x}},\mathbf{x},t) = J_W(\tilde{q},\tilde{p},q,p,t)$ is defined by the relation
\begin{equation}\label{JMap}
 W(\tilde {\mathbf x},t) = \int_{{\mathbb{R}^2}} J_W(\tilde {\mathbf x}, \mathbf x, t) W(\mathbf x,0) \, \rmd \mathbf x  \;,
\end{equation}
where we write $W(\mathbf{x},t) = W(q,p,t)$ with 
$\mathbf{x} = (q, p)^T$ and $\rmd \mathbf x = \rmd q \, \rmd p$ for abbreviation.
Note that $W(\mathbf x,t)$ and $J_W(\tilde {\mathbf x}, \mathbf x,t)$ are real functions.

A Gaussian state to given $\mathbf X(t)$, $\mathbf \Sigma(t)$ has the Wigner function
\begin{equation}\label{GaussianWigner}
 W_g(\mathbf{x},t) = \frac{\exp \big[ - \frac{1}{2} (\mathbf{x} - \mathbf{X}(t))  \cdot \mathbf{\Sigma}^{-1}(t) (\mathbf{x} - \mathbf{X}(t))
    \big] }{2 \pi \sqrt{\det \mathbf{\Sigma}(t)}} \;, 
 \end{equation}
and the most general expression for $J_W(\cdot)$ that respects this structure is an exponential function of the 14 linear and quadratic terms in the coordinates $q,p, \tilde{q}, \tilde{p}$. 
The normalization $\int_{\mathbb{R}^2} W(\mathbf x,t) \rmd\mathbf x =1 $ of Wigner functions implies the condition
\begin{equation}\label{JNorm}
 \int_{\mathbb{R}^2} J_W(\tilde {\mathbf x}, \mathbf x,t) \, \rmd \tilde{ \mathbf x} = 1 
\end{equation}
on the propagating function, which fixes the prefactors of the 5 terms $q^2, p^2, qp, q, p$ in the initial coordinates.
This leaves $9$ free parameters that have to be fixed in accordance with the linear transformations
~\eqref{X},~\eqref{Sigma} of expectation values.
The final result is
\begin{widetext}
\begin{equation}\label{JW}
 J_W(\tilde{\mathbf{x}},\mathbf{x},t) =  \frac{\exp \Big[ - \dfrac{1}{2} \big( \tilde{\mathbf{x}} - \mathbf{U}(t) \mathbf{x} - \mathbf{I}(t) \big) \cdot \mathbf{C}^{-1}(t) 
   \big( \tilde{\mathbf{x}} -  \mathbf{U}(t) \mathbf{x} - \mathbf{I}(t) \big) \Big] }
  {2 \pi \sqrt {\det \mathbf{C}(t)}} \;,
 \end{equation}
\end{widetext}
 where the $4+3+2=9$ parameters are the entries of the $2 \times 2$ matrix $\mathbf{U}(t)$ from Eq.~\eqref{UMat}, the symmetric and positive definite $2 \times 2$ matrix $\mathbf{C}$(t) from Eq.~\eqref{C}, and the two-dimensional vector $\mathbf{I}(t)$ from Eq.~\eqref{I}. 
 
 In order to check that this expression indeed reproduces the transformations~\eqref{X},~\eqref{Sigma}, we can express the expectation values at $t \ge 0$ in terms of those at $t=0$ through the evaluation of simple Gaussian integrals.
To give an example, it is
\begin{eqnarray}\label{QTrans1}
{\langle Q(t) \rangle} &=& \int_{\mathbb{R}^2} \tilde{q} \, W(\tilde {\mathbf x},t) \, \rmd \tilde{ \mathbf x} \nonumber\\
&=& \int_{\mathbb{R}^4} \tilde{q} J_W(\tilde {\mathbf x}, \mathbf x,t) W(\mathbf x,0)  \, \rmd \tilde {\mathbf x } \, \rmd \mathbf x \;.
\end{eqnarray}
The integral of $\tilde{q} J_W(\tilde {\mathbf x}, \mathbf x,t)$
over $\tilde {\mathbf x}$ is a Gaussian integral with a linear term,
and gives 
\begin{equation}\label{QTrans2}
  \int_{\mathbb{R}^2} \tilde{q} J_W(\tilde { \mathbf{x}}, \mathbf x,t) \, \rmd \tilde {\mathbf x} =
  U_{QQ}(t) q + U_{QP}(t) p + I_Q(t) \;.
\end{equation}
The final integration over $\mathbf x$ in Eq.~\eqref{QTrans1},
which now involves the right hand side of~\eqref{QTrans2}, 
generates the initial expectation values $\langle Q(0) \rangle$, $\langle P(0) \rangle$.
Therefore, we obtain the relation
$\langle Q(t) \rangle = U_{QQ}(t) \langle Q(0) \rangle + U_{QP}(t) \langle P(0) \rangle + I_Q(t) = \dot{u}(t) \langle Q(0) \rangle + u(t)  \langle P(0) \rangle + I_Q(t)$ in accordance with Eq.~\eqref{X}.
Following this recipe, we find that the given expression~\eqref{JW} for the propagating function $J_W(\cdot)$ reproduces the entire transformations~\eqref{X},~\eqref{Sigma} of the expectation values $\mathbf X(t)$, $\mathbf \Sigma(t)$, as we required.

If $\mathbf{C}(t) \to 0$,  we get a representation of the distribution $\delta(\tilde{\mathbf{x}}-\mathbf{U}(t) \mathbf{x} - \mathbf{I}(t))$ from Eq.~\eqref{JW}.
In particular for $t=0$,
where $\mathbf{U}(0) = 1$, $\mathbf{I}(0)=0$ in addition to $\mathbf{C}(0)=0$,
we have the correct result $J_W(\tilde{\mathbf{x}},\mathbf{x},0)=\delta(\tilde{\mathbf{x}}-\mathbf{x})$ in Eq.~\eqref{JMap}.

We note that the conveniently simple derivation of $J_W(\cdot)$ relies on the use of Wigner functions.
Of course, the expressions for $\rho_S(q_f, q_f',t)$ in position representation often reported in the literature can be recovered from Eq.~\eqref{JW} (see App.~\ref{app:Prop}).

\section{\label{sec:Equil}Equilibration and thermalization}

The results from the previous section allow us to study the behavior of the central oscillator density matrix $\rho_S(t)$ in the long-time limit $t \to \infty$.
We can classify the behavior according to the general criteria of equilibration and thermalization.
Equilibration means convergence to a stationary state
as expressed in the two conditions
\begin{itemize}
\item[(E1)] the central oscillator density matrix $\rho_S(t)$ converges for $t \to \infty$,
\item[(E2)] the stationary state $\rho_S^\infty=\lim_{t \to \infty} \rho_S(t)$ is independent of $\rho_S(0)$ \;.
\end{itemize}
Note that $\rho^\infty_S$ will depend on the initial bath state $\rho_B(0)$.
Note further that the above definition of equilibration does not distinguish between stationary equilibrium states and stationary non-equilibrium states with finite heat flows. The latter cannot arise for a single bath with continuous initial conditions as in Eq.~\eqref{SigmaAnsatz} such that condition (E1) is sufficient for the present study.

Equilibration (E1) implies convergence of central oscillator expectation values for $t \to \infty$.
This, in turn, requires convergence of the matrix $\mathbf U(t)$ in Eqs.~\eqref{X},~\eqref{Sigma}.
Because the only stationary solution of the homogeneous differential Eq.~\eqref{uhom} is $u(t) \equiv 0$,
convergence of $\mathbf U(t)$ is equivalent to $\mathbf U(t) \to 0$ or $u(t) \to 0$ for $t \to \infty$.
Therefore, we assume in this section the condition
\begin{itemize}
\item[(E0)]  $u(t) \to 0$ for $t \to \infty$
\end{itemize}
 as the prerequisite for equilibration (E1).
Under this assumption, we will be able to show convergence of expectation values and, building on this result, convergence of the central oscillator density matrix.

In the weak damping limit, 
condition (E0) is equivalent to $\gamma(\Omega)>0$ (taking the thermodynamic limit for granted).
This expresses the basic fact that equilibration occurs through energy exchange with the environment, which is not possible for an isolated oscillator with $\gamma(\Omega)=0$.
We note that a small value of $\gamma(\Omega)$ can result in long transients 
that prevent equilibration over the observation time.

Thermalization additionally requires that the stationary state $\rho_S^\infty$ is a thermal state,
and we have the three increasingly stronger properties
\begin{itemize}
\item[(T1)] the stationary state $\rho_S^\infty$ is a thermal state, 
\item[(T2)] the stationary state is a thermal state $\rho_S^\infty \propto \rme^{-H_S/T_\infty}$ of the central oscillator,
\item[(T3)] the temperature $T_\infty$ of the stationary thermal state $\rho_S^\infty$ is independent of the central oscillator frequency.
\end{itemize}
We will see that the stationary state is always Gaussian, which implies property (T1). Property (T2) reduces to an equipartition condition on the central oscillator variances that determine the Gaussian state, while property (T3) leads to a strong condition on the initial bath state.

\subsection{Expectation values in the long-time limit}

The assumption $\mathbf U(t) \to 0$  for $t \to \infty$ implies that
the terms $\mathbf{U}(t) \mathbf{X}(0)$ in Eq.~\eqref{X}
and $\mathbf{U}(t) \mathbf{\Sigma}(0) \mathbf{U}^T(t)$ in Eq.~\eqref{Sigma}
drop out of the expressions for $\mathbf{X}(t)$ and $\mathbf\Sigma(t)$ in the long-time limit.
Only the terms $\mathbf{I}(t)$ and $\mathbf{C}(t)$, which depend exclusively on the initial bath preparation, can survive the $t \to \infty$ limit:
All information about the initial central oscillator state is lost.
We can not immediately draw a conclusion about the long-time behavior because the functions $\tilde{u}(t,\omega)$, $\tilde{v}(t,\omega)$ from Eqs.~\eqref{UFour},~\eqref{UFourDot} do not converge for $t\to \infty$.
Instead, we note that $\tilde{u}(t,\omega)$ behaves asymptotically as
\begin{equation}\label{UAsym}
  \tilde{u}_{as}(t,\omega) \simeq \rme^{\rmi \omega t} \int_0^\infty u(\tau) \, \rme^{-\rmi \omega \tau} \, \rmd \tau  \qquad (t \to \infty) \;.
\end{equation}
Similarly, it follows $\tilde{v}(t,\omega) \simeq \rmi \omega \tilde{u}_{as}(t,\omega)$ for $t \to \infty$ from Eq.~\eqref{UFourDot}.
Consequently, the matrix $\mathbf{U}(t,\omega)$ behaves asymptotically as
\begin{equation}
 {\mathbf U}(t, \omega) \simeq  \begin{pmatrix} \Re \tilde{u}_{as}(t,\omega) & \dfrac{ \Im \tilde{u}_{as}(t,\omega)}{\omega} \\[2ex] -\omega \Im \tilde{u}_{as}(t,\omega) & \Re \tilde{u}_{as}(t,\omega) \end{pmatrix}  \quad (t \to \infty) \;,
\end{equation}
and remains oscillating for $t \to \infty$ even if $u(t) \to 0$.

The contributions to the term $\mathbf{I}(t)$ in Eq.~\eqref{ICont}, say to $\langle Q(t)\rangle$, are of the form 
 \begin{equation}
- \Re \int_0^\infty D(\omega) \lambda(\omega) \tilde{u}_{as}(t,\omega) \breve{X}_Q(\omega) \, \rmd \omega \;.
 \end{equation}
The integrand depends on $t$ through the factor $\rme^{\rmi \omega t}$ from Eq.~\eqref{UAsym},
such that the integral is the Fourier  transform
of an integrable (by assumption even continuous) function of $\omega$.
If we recall the Riemann-Lebesgue lemma~\eqref{ftprop}
we see that $\mathbf{I}(t) \to 0$ for $t \to \infty$.
Altogether, it follows that
the position and momentum expectation values vanish
in the long-time limit, i.e.  $\mathbf{X}(t) \to 0$ for $t \to \infty$.

For the variances,
a finite contribution can survive the $t \to \infty$ limit
because the squares of the matrix elements of $\mathbf{U}(t,\omega)$ occur in $\mathbf{C}(t)$.
For example, the diagonal term $\mathbf{C}^{(1)}(t)$ from Eq.~\eqref{CCont} contributes to $\Sigma_{QQ}(t)$ the integral
\begin{equation}\label{C1Explicit}
 C^{(1)}_{QQ}(t) = \int_0^\infty \omega \gamma(\omega) \, c_{QQ}(t,\omega) \, \rmd \omega
\end{equation}
of the function
\begin{eqnarray}\label{C1QQ}
  c_{QQ}(t,\omega) &=& [\Re \tilde{u}(t,\omega)]^2 \, \breve{\Sigma}_{QQ}^{(1)}(\omega) \nonumber\\
  &&+ \frac{2 [\Re \tilde{u}(t,\omega)] [\Im \tilde{u}(t,\omega)] }{\omega} \breve{\Sigma}_{QP}^{(1)}(\omega) \nonumber\\
  &&+ \frac{[\Im \tilde{u}(t,\omega)]^2}{\omega^2} \breve{\Sigma}_{PP}^{(1)}(\omega) \;.
\end{eqnarray}

Here we write, using the notation from Eq.~\eqref{SigmaMatrix},
\begin{equation}
 {\mathbf C}^{(1)}(t) = \begin{pmatrix}  C^{(1)}_{Q Q}(t) &  C_{Q P}^{(1)}(t) \\[1ex]  C_{Q P}^{(1)}(t) &  C_{P P}^{(1)}(t) \end{pmatrix}
\end{equation}
for the matrix elements of $ {\mathbf C}^{(1)}(t)$ and
\begin{equation}
 \breve{\mathbf \Sigma}^{(1)}(\omega) = \begin{pmatrix} \breve \Sigma^{(1)}_{Q Q}(\omega) & \breve \Sigma_{Q P}^{(1)}(\omega) \\[1ex] \breve \Sigma_{Q P}^{(1)}(\omega) & \breve \Sigma_{P P}^{(1)}(\omega) \end{pmatrix} 
\end{equation}
for the matrix elements of $ \breve{\mathbf \Sigma}^{(1)}(\omega)$ from Eq.~\eqref{SigmaAnsatz}.

The contribution from the first term in $c_{QQ}(t,\omega)$ is
\begin{equation}
 \int_0^\infty \omega \gamma(\omega) [\Re \tilde{u}_{as}(t,\omega)]^2 \breve\Sigma^{(1)}_{QQ}(\omega) \, \rmd\omega \;.
\end{equation}
If we expand the square $ [\Re \tilde{u}_{as}(t,\omega)]^2$ according to
\begin{equation}\label{exprcompl}
 [\Re \rme^{\rmi \omega t} z]^2 = \frac{|z|^2}{2} + \frac{z_r^2-z_i^2}{2} \cos 2 \omega t - z_r z_i \sin 2 \omega t \,,
\end{equation}
for a complex number $z$ with $z_r = \Re z$, $z_i = \Im z$,
we see that in the above integral a contribution $|\tilde{u}_{as}(t,\omega)|^2/2$ remains finite for $t \to \infty$, while the oscillatory terms with $\cos 2 \omega t$, $\sin 2 \omega t$ vanish according to the Riemann-Lebesgue lemma~\eqref{ftprop}.
Similar expressions are obtained for the remaining terms in $\mathbf{C}^{(1)}(t)$.

The off-diagonal term $\mathbf{C}^{(2)}(t)$ from Eq.~\eqref{CContOD} is given by a double Fourier integral
and contains only oscillatory terms in the two frequencies $\omega_1$, $\omega_2$.
Therefore, $\mathbf{C}^{(2)}(t) \to 0$ for $t \to \infty$. 

We can now collect the finite contributions from the different terms in $\mathbf{C}^{(1)}(t)$, to find that the central oscillator variances converge to stationary values 
$\mathbf{\Sigma}^\infty = \lim_{t \to \infty} \mathbf\Sigma(t)$ 
in the long-time limit. 
They are given by
\begin{equation}\label{CQQLong}
 \Sigma^\infty_{QQ} =  \int_0^\infty \gamma(\omega)   \left|\int_0^\infty \rme^{\rmi \tau \omega} u(\tau) \, \rmd\tau \right|^2 \frac{\breve{\mathcal{E}}(\omega)}{\omega} \, \rmd \omega \,, 
 \end{equation}
 \begin{equation}\label{CPPLong}
 \Sigma^\infty_{PP} =  \int_0^\infty \gamma(\omega)   \left|\int_0^\infty \rme^{\rmi \tau \omega} u(\tau) \, \rmd\tau \right|^2 \omega \breve{\mathcal{E}}(\omega) \, \rmd \omega \,, 
 \end{equation}
 \begin{equation}\label{CQPLong}
 \Sigma^\infty_{QP} =  0 \,,
\end{equation}
where 
\begin{equation}\label{ECal}
 \breve{\mathcal{E}}(\omega) = \frac{1}{2} \left(\omega^2 \breve{\Sigma}^{(1)}_{QQ}(\omega) + \breve{\Sigma}_{PP}^{(1)}(\omega) \right) \;.
\end{equation}
Comparison with Eqs.~\eqref{UFromFourier2},~\eqref{FAnalytic}
gives the alternative expressions
\begin{equation}\label{CQQLongF}
 \Sigma^\infty_{QQ} =   \frac{2}{\pi} \int_0^\infty  \Im F(\omega + \rmi 0^+) \,\frac{\breve{\mathcal{E}}(\omega)}{\omega} \, \rmd\omega \,, 
 \end{equation}
 \begin{equation}\label{CPPLongF}
 \Sigma^\infty_{PP} =  \frac{2}{\pi} \int_0^\infty \Im F(\omega + \rmi 0^+) \, \omega \breve{\mathcal{E}}(\omega) \, \rmd\omega \,.
 \end{equation}
Recall that $F(\omega+\rmi 0^+)$ is a continuous function according to our assumption $u(t) \to 0$.

As noted before, the values $\mathbf{\Sigma}^\infty$ are independent of the initial central oscillator state. Furthermore, the initial bath state $\rho_B(0)$ occurs only through the frequency-resolved
energy distribution $\breve{\mathcal{E}}(\omega)$.
In particular, 
the known equations for thermal baths~\cite{HR85}
 are recovered whenever $\breve{\mathcal{E}}(\omega) = E(T,\omega)$,
where
\begin{equation}\label{thermprop}
E(T,\Omega) =  \frac{\Omega}{2} \coth \frac{\Omega}{2T} 
\end{equation}
is the energy of a thermal oscillator at temperature $T$.
 Because there are no separate conditions
on the two functions $\breve{\Sigma}^{(1)}_{QQ}(\omega)$, $\breve{\Sigma}^{(1)}_{PP}(\omega)$,
thermalization is possible also in non-thermal environments (see below).

Eqs.~\eqref{CQQLong}--\eqref{CQPLong} follow directly if we assume a thermal bath from the outset, with initial conditions 
$\omega^2 \breve{\Sigma}^{(1)}_{QQ} (\omega) = \breve{\Sigma}^{(1)}_{PP} (\omega) = E(T,\omega)$ and $\breve{\Sigma}^{(1)}_{QP}(\omega)=0$.
Equipartition of energy 
allows us to combine the terms in Eq.~\eqref{C1QQ} to $c_{QQ}(t,\omega)=|\tilde u (t,\omega)|^2 E(T,\omega)/\omega^2$, which depends only on the modulus of $\tilde{u}(t,\omega)$.
We can then drop the exponential factor $\rme^{\rmi \omega t}$ from Eq.~\eqref{UAsym},
and convergence of $\mathbf \Sigma(t)$ is evident. This short cut is not available in the general case.

\subsection{Equilibration of the central oscillator}

If the initial bath state $\rho_B(0)$ and the central oscillator state $\rho_S(0)$ are both Gaussian states,
the central oscillator density matrix $\rho_S(t)$ is Gaussian for all $t \ge 0$.
Then, $\rho_S(t)$ is completely determined by the values of $\mathbf X(t)$, $\mathbf \Sigma(t)$, and their convergence suffices to establish equilibration (E1), and also (E2), in this case.

Otherwise, for non-Gaussian initial states $\rho_S(0)$, 
we can use the propagating function
$J_W(\tilde{\mathbf{x}},\mathbf{x},t)$ from Eq.~\eqref{JW} to find $\rho_S(t)$ for $t \to \infty$.
Recall that according to Sec.~\ref{sec:TDLimit} we can assume that the initial bath state 
is Gaussian in the thermodynamic limit, which allows for the construction given in Sec.~\ref{sec:prop}.

Equilibration follows now from the observation that $J_W(\tilde{\mathbf{x}},\mathbf{x},t)$ 
 converges for $t \to \infty$ whenever $\mathbf X(t)$, $\mathbf \Sigma(t)$ converge.
The long-time limit
\begin{equation}\label{JWLong}
 J^\infty_W(\tilde{\mathbf{x}}) = \lim_{t \to \infty} J_W(\tilde{\mathbf{x}},\mathbf{x},t) = 
 \frac{\exp \Big[\!  - \dfrac{1}{2} \tilde{\mathbf{x}}  \cdot (\mathbf{\Sigma}^\infty)^{-1} 
   \tilde{\mathbf{x}}  \Big] }
  {2 \pi \sqrt {\det \mathbf{\Sigma}^\infty}} \;.
\end{equation}
is obtained
 through substitution of $\lim_{t\to\infty} \mathbf I(t)= 0$ and 
$\lim_{t\to\infty} \mathbf C(t)= \mathbf \Sigma^\infty$ from Eqs.~\eqref{CQQLong}---\eqref{CQPLong}.
Because $\mathbf U (t) \to 0$, the result 
does not depend on $\mathbf{x}$.

The long-time limit of the Wigner function $W_S^\infty(\mathbf x)=\lim_{t \to \infty} W_S(\mathbf x,t)$ follows immediately with Eq.~\eqref{JMap}:
The integration over $\mathbf x$ in the resulting expression
\begin{equation}
 W_S^\infty(\tilde { \mathbf x}) = \int_{\mathbb{R}^2} J_W^\infty(\tilde {\mathbf x}) W_S(\mathbf x,0) \, \rmd \mathbf x  = J_W^\infty(\tilde {\mathbf x})  
\end{equation}
evaluates to one because $W_S(\mathbf x,0)$ is normalized,
such that $W_S^\infty(\mathbf x)$ is equal to $J^\infty_W(\mathbf{x})$.
In other words, the stationary state $\rho_S^\infty$ is a Gaussian state~\eqref{GaussianWigner} with parameters $\mathbf X = 0$, $\mathbf{\Sigma} = \mathbf{\Sigma}^\infty$.
These parameters 
depend on the initial bath state according to Eqs.~\eqref{CQQLong}, \eqref{CPPLong}, but they are independent from the initial central oscillator state. 
This proves equilibration (E1) and (E2) for general initial central oscillator states. 
In particular, the stationary state is Gaussian  also for non-Gaussian initial states.

We note that the propagating function in position representation does not converge in the long-time limit (cf. App.~\ref{app:Prop}), which prevents an equally simple argument.

\subsection{Thermalization of the central oscillator}

Because the stationary state $\rho_S^\infty$ in the long-time limit is a Gaussian state for which only $\Sigma^\infty_{QQ}$, $\Sigma^\infty_{PP}$ are non-zero,
it can always be interpreted as the thermal equilibrium state of some harmonic oscillator.
This establishes the weakest thermalization property (T1).

The effective oscillator frequency $\Omega_\infty$ and temperature $T_\infty$ associated with 
$\rho_S^\infty$
are
\begin{equation}\label{OmTLong}
\Omega^2_\infty = \frac{\Sigma^\infty_{PP}}{\Sigma^\infty_{QQ}} \;, \qquad T_\infty = \frac{\Omega_\infty}{2} \mathrm{arcoth}^{-1} \Big[ 2 \sqrt{ \Sigma^\infty_{QQ} \Sigma^\infty_{PP} } \, \Big] \,.
\end{equation}
Generally, $\Omega_\infty$  
is not equal to the central oscillator frequency $\Omega$
such that the stronger property (T2) is not fulfilled.
By Eq.~\eqref{OmTLong}, the condition $\Omega_\infty = \Omega$ is equivalent to 
equipartition of kinetic and potential energy $ \langle P^2\rangle = \Sigma^\infty_{PP} = \Omega^2 \Sigma^\infty_{QQ} =\Omega^2 \langle Q^2\rangle$.
The violation of this condition arises from
the integrations over $\omega$ in Eqs.~\eqref{CQQLong},~\eqref{CPPLong} or~\eqref{CQQLongF},~\eqref{CPPLongF}, which cover a finite energy range and include values $\omega \ne \Omega$.
Quantum corrections of this type are characteristic for strong damping~\cite{HR85}.

Equipartition of energy is achieved in the limit of weak damping ($\gamma(\Omega) \to 0$, when according to Eq.~\eqref{FAnalytic}
the function $(2/\pi) \Im F(\omega + \rmi 0^+)$ in Eqs.~\eqref{CQQLongF},~\eqref{CPPLongF} 
converges to $2 \delta(\omega^2 - \Omega^2)= (\delta(\omega+ \Omega) + \delta(\omega-\Omega))/\Omega$.
Therefore, the values
$\Omega^2 \Sigma_{QQ}^\infty = \Sigma_{PP}^\infty = \breve{\mathcal{E}}(\Omega)$ 
are obtained.
This establishes the stronger thermalization property (T2) in the weak damping limit.

For these values of $ \Sigma_{QQ}^\infty, \Sigma_{PP}^\infty$ it is (cf. Eq.~\eqref{thermprop})
\begin{equation}\label{TWC}
 \Omega^{(\mathrm{WD})}_\infty = \Omega \;, \qquad T^{(\mathrm{WD})}_\infty(\Omega) = \frac{\Omega}{2} \mathrm{arcoth}^{-1} \, \frac{2 \breve{\mathcal{E}}(\Omega)}{\Omega} \,,
\end{equation}
such that the stationary state $\rho^\infty_S$ is a thermal equilibrium state of the central oscillator.
The temperature $T_\infty(\Omega)$ is determined by the energy $\breve{\mathcal{E}}(\Omega)$ of the bath oscillator at frequency $\Omega$ in the initial state.
Note that the assumption (E0) implies $\gamma(\Omega) \ne 0$ and $D(\omega) \ne 0$, such that the value of $\breve{\mathcal{E}}(\Omega)$ is defined.
In particular, $\breve{\mathcal{E}}(\Omega) \ge \Omega/2$ and the argument of $\mathrm{arcoth}(\cdot)$ is equal to or greater than one.

Still, the asymptotic temperature $T_\infty=T^{(\mathrm{WD})}_\infty(\Omega)$ from Eq.~\eqref{TWC}
is a function of $\Omega$.
The functional dependence is determined by the choice of $\breve{\mathcal{E}}(\omega)$.
If we demand, for the strongest thermalization property (T3), that $T_\infty$ is independent of $\Omega$
we have to solve Eq.~\eqref{TWC} 
to obtain the condition
\begin{equation}\label{ThermAtWC}
\breve{\mathcal{E}}(\omega)  = \frac{\omega}{2} \coth \frac{\omega}{2T_\infty} \;.
\end{equation}
Note that this is a condition on the particular combination $\breve{\mathcal{E}}(\omega)$ of initial bath variances $\breve{\Sigma}^{(1)}_{QQ}(\omega)$, $\breve{\Sigma}^{(1)}_{PP}(\omega)$, and not on the individual functions.
Therefore, any initial bath preparation with $  \breve{\mathcal{E}}(\omega) = E(T_0,\omega)$ results in the same stationary states as the thermal bath at temperature $T_0$.
One example for this additional freedom is the choice
\begin{equation}\label{NonThermalThermalization}
 \breve{\Sigma}^{(1)}_{QQ}(\omega)= \frac{\coth (\omega/2 T_0)  - 1/2}{\omega}  \;, \qquad \breve{\Sigma}_{PP}^{(1)}(\omega)  = \omega/2  \;,
\end{equation}
and arbitrary $\breve{\Sigma}_{QP}^{(1)}(\omega)$.
It can be realized, e.g., by superposition of coherent oscillator states at different positions. 
This initial bath state is not a thermal  state for $T_0 > 0$, in particular it violates equipartition of energy $\omega^2 \breve{\Sigma}^{(1)}_{QQ}(\omega) = \breve{\Sigma}_{PP}^{(1)}(\omega)$.
But since $\breve{\mathcal{E}}(\omega) = E(T_0,\omega)$ we find that the stationary central oscillator state $\rho_S^\infty$ is identical to that obtained with a thermal bath at temperature $T_0$:
Thermalization is well possible in non-thermal environments, even those far from thermal equilibrium.

\subsection{\label{sec:EquilWithList}Summary}
In summary, we have a hierarchy of conditions for equilibration and thermalization:
\begin{enumerate}\itemindent6ex\hangindent6ex
\item[(E1), (E2)] the central oscillator equilibrates whenever $u(t) \to 0$ for $t \to \infty$,
\item[(T1)] the stationary state is  always a Gaussian and thermal state,
\item[(T2)] equipartition of kinetic and potential energy occurs precisely at weak damping,
\item[(T3)] the asymptotic temperature $T_\infty$ is independent of the central oscillator frequency under the additional condition~\eqref{ThermAtWC} on $\breve{\mathcal{E}}(\omega)$.
\end{enumerate}
It is a special feature of linear systems such as the one studied here
that equilibration depends only on the asymptotic behavior of the solution $u(t)$ of a classical equation of motion~\eqref{uhom}.
Another feature is that the stationary state always is Gaussian such that equilibration implies thermalization, albeit only in the weak sense of property (T1).
We noted earlier that in the situation studied here, with coupling to a single bath, a stationary state does not admit finite heat flows as would become possible for several baths with different preparations $\breve {\mathcal E}(\omega)$. Therefore, conditions (E1), (E2) capture the standard notion of thermodynamic equilibrium.

We note that a consistent definition of thermalization requires the strong property (T3).
Suppose we deal with two central oscillators with frequencies $\Omega_1 \ne \Omega_2$.
In the weak damping limit, the stationary state is the product state of two independent thermal states with respective temperatures $T_\infty(\Omega_1)$ and $T_\infty(\Omega_2)$. 
Such a state is only a thermal state of the combined system comprising the two oscillators 
if $T_\infty(\Omega_1)=T_\infty(\Omega_2)$.
Therefore, thermalization of multiple oscillators, already in the weak sense (T1), requires the strong property (T3) and thus condition~\eqref{ThermAtWC}
(but recall that this condition can be fulfilled also for non-thermal environments as in Eq.~\eqref{NonThermalThermalization}).

\section{\label{sec:Chain}The infinite harmonic chain}

\begin{figure}
\includegraphics[width=\linewidth]{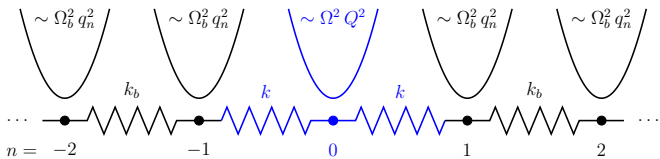}
\caption{\label{fig:Chain}(Color online) Sketch of the infinite harmonic chain as defined in Eqs.~\eqref{ChainHB},~\eqref{ChainHSB}.}
\end{figure}

As an example for equilibration in a non-thermal environment
we consider an infinite chain of harmonic oscillators (see Fig.~\ref{fig:Chain}).
Oscillators in the right ($n \ge1 $)  and left ($n \le -1$) half of the chain, with frequency $\Omega_b$, are coupled to their neighbors ($n \pm 1$) with spring constant $k_b$.
They form the harmonic oscillator bath for the central oscillator at $n=0$,
with oscillator frequency $\Omega$ and coupling $\kc$ to the oscillators at $n=\pm 1$.
For $\Omega=\Omega_b$ and $\kc=k_b$ we obtain a homogeneous, translational invariant chain.

Related examples have been studied in numerous publications, see e.g.~\cite{Ull66c,Rub60,RH69,Aga71,HRN71,TS94,TY94,DR09}.
The behavior for thermal initial conditions, e.g. in a homogeneous chain~\cite{Rub60} or a chain with a single heavy mass~\cite{Ull66c}, is well understood.
Equilibration in a harmonic chain with non-thermal initial conditions as discussed in Refs.~\cite{RH69,Aga71} can be expressed in terms of our conditions from Sec.~\ref{sec:EquilWithList}.
General arguments for the appearance of Gaussian states in the long-time limit are given in~\cite{TS94,CE10}.
Still, a satisfactory and explicit analysis of equilibration and thermalization of the simple chain in non-thermal environments is missing.
Some studies assume too quickly that equilibration implies thermalization, in the sense of our condition (T1), failing to note, e.g., that the appearance of Gaussian states  is the general behavior of linear systems and unrelated to thermalization as expressed by condition (T3).
According definitions of `temperature' have to be taken with care.
In addition we must carefully analyze the role of undamped oscillatory behavior that prevents equilibration and, therefore, thermalization. 

\subsection{Mapping onto the central oscillator model}

To address the harmonic chain within the formalism from Secs.~\ref{sec:model}---\ref{sec:Equil} we must transform the Hamilton operator
\begin{equation}\label{ChainHB}
 H_B = \frac{1}{2} \sum_{n=1}^\infty  \Big[ p_n^2 + \Omega_b^2 q_n^2 \Big]   -  k_b \sum_{n=1}^\infty q_n q_{n+1}
\end{equation}
for the harmonic oscillator bath (with operators $q_n$, $p_n$ for the oscillator at site $n \ne 0$) to normal modes.
The same transformation has to be applied to the operator 
$k q_1$ in the coupling term
\begin{equation}\label{ChainHSB}
H_{SB} = - \kc \, Q ( q_1 + q_{-1} )
\end{equation}
between the central oscillator and the chain oscillators at $n = \pm 1$.
It suffices to treat one of the two half-infinite chains explicitly, say the right chain $n \ge 1$ as in Eq.~\eqref{ChainHB},
and include a factor of two in $\gamma(\omega)$ to account for the left chain $n \le -1$.
Note that in doing so we implicitly assume identical initial conditions for both sides of the chain and thus exclude the possibility of stationary non-equilibrium states with finite heat flow between the right and left half-infinite chain. 

The normal modes of $H_B$ are the standing wave solutions
$f_\nu(n) \propto \sin \big( \frac{\pi \nu n}{N+1} \big)$ (for a finite chain of length $N$),
and after a few lines of calculation we obtain the spectral function
\begin{equation}\label{SpecChain}
 \gamma(\omega) = \frac{2}{\pi} \frac{\kc^2}{k_b^2}  \sqrt{4 k_b^2 - \big( \Omega_b^2 - \omega^2 \big)^2}  \qquad \mathrm{for} \;  |\Omega_b^2-\omega^2| < 2 k_b 
\end{equation}
in the thermodynamic limit $N \to \infty$.
It is $\gamma(\omega)=0$ for $|\Omega_b^2-\omega^2| > 2 k_b$,
and we impose the positivity condition $\Omega_b^2 \ge  2 k_b \ge 0$
to exclude negative frequencies of the bath.

To proceed it is convenient to introduce the dimensionless model parameters
\begin{equation}\label{NormalizedParams}
  \kappa_b = \frac{2 k_b}{\Omega_b^2} \;, \qquad
  \kac = \frac{2 \kc}{\Omega_b^2} \;, \qquad
  \Omega_r = \frac{\Omega}{\Omega_b} \;,
\end{equation}
and to use the normalized quantities
\begin{equation}
  \bar{\omega} = \frac{\omega}{\Omega_b} \;, \qquad
  \bar{t} = t \Omega_b \;, \qquad
  \bar{u}(\bar{t}) =  \Omega_b u(\bar{t}) \;.
\end{equation}
Note that $0 \le \kappa_b \le 1$.

\subsection{Conditions for equilibration in the harmonic chain}

As discussed in Sec.~\ref{sec:Equil}, equilibration depends entirely on the decay of the function $u(t)$ for $t \to \infty$, and thus on the absence of poles in $F(z)$ from Eq.~\eqref{FCont}.
To obtain $F(\omega)$, we use the representation~\eqref{FAnalytic} with the complex function
\begin{equation}
 \Gamma(z) = \frac{\kc^2}{k_b^2} \Big( z^2 - \Omega_b^2 \mp \sqrt{(\Omega_b^2 - z^2)^2 - 4 k_b^2}  \, \Big) \;,
\end{equation}
where the branch cut of the root must be chosen along the positive real axis,
and the minus (plus) sign applies for $\Re z > 0$ ($\Re z < 0$).
Note that the positivity condition~\eqref{pos}, which can now be rewritten as $\Omega^2 +\Gamma(\rmi 0^+) \ge 0$, requires that
\begin{equation}\label{PosForChain}
 \Omega_r^2 \ge \frac{\kappa^2}{\kappa_b^2} \left( 1 - \sqrt{1 -  \kappa_b^2} \right)\;.
\end{equation}

\begin{figure}
\includegraphics[width=0.7\linewidth]{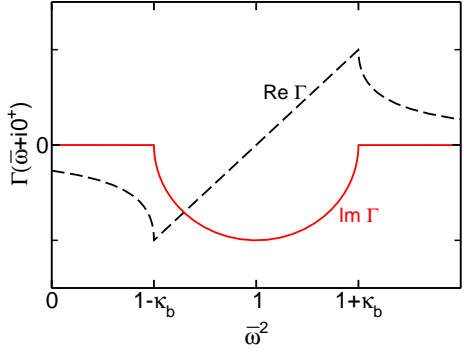}
\caption{\label{fig:GamZ} (Color online) Real (dashed curve) and imaginary (solid curve) part of $\Gamma(\omega + \rmi 0^+)$ for $\kappa_b=1/2$ and $\omega > 0$.
For $\bar{\omega}^2 = \{ 1-\kappa_b , 1 , 1+\kappa_b \}$ the function value is 
$\{-1, -\rmi, 1\} \times (\kac \Omega_b)^2/\kappa_b$, respectively.}
\end{figure}

Before we can determine the function $u(t)$ with Eq.~\eqref{UFromFourier2}
we must consider the possibility of isolated poles of $F(z)$.
According to Eq.~\eqref{FAnalytic} we have to compare the functions $\omega^2 - \Omega^2$ and $\Re \Gamma(\omega + \rmi 0^+)$ in regions where $\Im \Gamma(\omega + \rmi 0^+) = 0$.
From the qualitative behavior of $\Gamma(\omega + \rmi 0^+)$, shown in Fig.~\ref{fig:GamZ}, 
we deduce that isolated poles of $F(z)$ do not exist if and only if 
the inequalities
\begin{equation}\label{NoPoles}
  1 - \frac{\kappa_b^2 - \kac^2}{\kappa_b} \le \Omega_r^2 \le 1 + \frac{\kappa_b^2 - \kac^2}{\kappa_b}
\end{equation}
are fulfilled.
The first inequality excludes poles in the interval $\bar{\omega}^2 < 1 - \kappa_b$,
the second inequality in the interval $\bar{\omega}^2 > 1 + \kappa_b$.
Another more fundamental restriction is the positivity condition~\eqref{PosForChain}, which is however less restrictive than the present condition.

\begin{figure}
\includegraphics[width=0.7\linewidth]{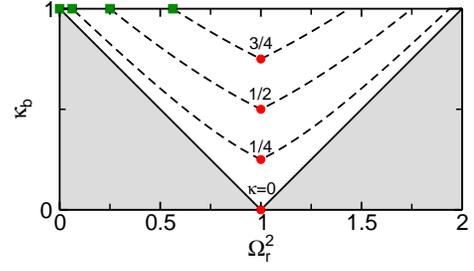}
\caption{\label{fig:params} (Color online) Diagram of the admissible parameter space for equilibration according to condition~\eqref{NoPoles}.
The white triangular region above the solid black lines is the maximal set of allowed parameter combinations.  
Outside of this region an isolated pole exists even in the weak damping limit $\kac \to 0$.
For $\kac > 0$, the region of admissible parameters shrinks as depicted by the dashed black curves.
The parameter combinations of homogeneous chains ($\Omega_r=1$) corresponds to the cusps $\kappa_b=\kappa$ of the curves, marked with red dots.
The parameter combinations of chains with a single heavy mass~\cite{Ull66c}
correspond to the intersections of the curves with the $\kappa_b=1$ line at $\Omega_r=\kappa$,
marked with green squares.
At these points, condition~\eqref{NoPoles} coincides with the positivity condition~\eqref{PosForChain}.}
\end{figure}

The admissible parameter combinations for equilibration of the harmonic chain that follow from condition~\eqref{NoPoles} are depicted in Fig.~\ref{fig:params}.
We note the basic restrictions
\begin{equation}\label{NoPolesWC}
\kac \le \kappa_b \qquad \rm{and} \qquad |1-\Omega_r^2| \le \kappa_b \;.
\end{equation}
The second inequality guarantees that the central oscillator frequency $\Omega_r$ lies within the interval $\bar{\omega} \in [\sqrt{1-\kappa_b},\sqrt{1+\kappa_b}]$ where $\gamma(\bar{\omega})>0$.
If this is fulfilled, equilibration is always possible for sufficiently small $\kac$.
Since $\kappa_b \le 1$, it restricts the admissible parameters to the rectangle $(\kappa_b, \Omega_r^2) \in [0,1]\times [0,2]$.

Condition~\eqref{NoPoles} is always fulfilled for the homogeneous chain
(and we note that $\kappa=\kappa_b$ requires $\Omega_r=1$).
The chain studied by Ullersma corresponds to parameters $\kappa_b=1$ and $\Omega_r = \kappa$ ($\Omega_r^2$ equals the mass ratio $\mu$ in~\cite{Ull66c}). 
Condition~\eqref{NoPoles} is fulfilled if $\Omega_r \le 1$, i.e. only for a heavy mass.
Both examples lie on the boundary of the admissible parameter space,
with one or two of the inequalities in~\eqref{NoPoles} becoming equalities.

\subsection{Dynamical evolution of the harmonic chain}

Depending on parameters, the harmonic chain features rich dynamical behavior.
For parameter combinations that fulfill condition~\eqref{NoPoles}
the explicit result for $u(t)$ from Eq.~\eqref{UFromFourier2} reads 
\begin{eqnarray}\label{UChain}
 \bar{u}(\bar{t}) &=& \frac{2  \kappa^2}{\pi}  \int_{\sqrt{1 - \kappa_b}}^{\sqrt{1+ \kappa_b}} \sin \bar{\omega} \bar{t} \, \sqrt{\kappa_b^2 - (1 - \bar{\omega}^2)^2}   \nonumber\\
  &&\times  \frac{1}{\kappa_b^2 (\bar{\omega}^2-\Omega_r^2)^2 - 2 \kappa^2 (\bar{\omega}^2-1)(\bar{\omega}^2-\Omega_r^2)
 + \kappa^4} \, \rmd\bar{\omega} \;. \nonumber\\
\end{eqnarray}
For parameter combinations violating condition~\eqref{NoPoles} 
isolated poles of $F(z)$ occur and additional (undamped) sine functions $\xi_i \sin \bar \Omega_i t$ must be added to this expression. According to Eq.~\eqref{FAnalytic}, the poles of $F(z)$ are the solutions of $\Omega^2 - \Omega_i^2 + \Gamma(\Omega_i) = 0$,
which gives a quadratic equation for the harmonic chain such that zero, one, or two (positive) poles are possible.

\begin{figure}
\includegraphics[width=0.8\linewidth]{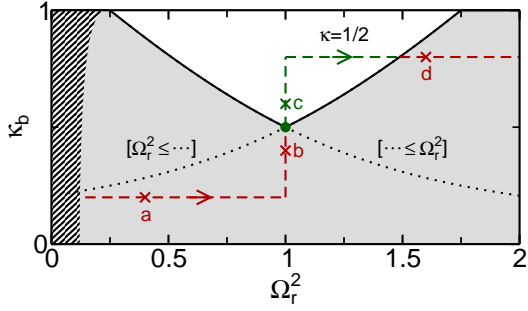}
\caption{\label{fig:paramsK2} (Color online) The $(\Omega_r^2,\kappa_b)$ parameter space of the infinite harmonic chain for $\kappa = 1/2$.
The solid/dotted black curves give the boundary of the two regions   
defined by each of the inequalities from condition~\eqref{NoPoles}.
The central oscillator equilibrates for parameters in the unshaded region above the solid curve, where the condition is fulfilled (note that $\kappa_b \le 1$).
For parameters lying between the solid and dotted curves one of the two inequalities is violated and a single isolated pole of $F(z)$ exists.
Below the dotted curve $F(z)$ has two isolated poles.
The dashed region to the left indicates where the positivity condition~\eqref{PosForChain} is violated, but such parameters already violate the first inequality in~\eqref{NoPoles}.
Both conditions coincide at  $\kappa_b=1$, $\Omega_r=\kappa$.
The  dashed red/green lines indicate the path followed in the next Fig.~\ref{fig:isomod},
the crosses marked a---d indicate the parameters used in Fig.~\ref{fig:ut}.
}
\end{figure}

\begin{figure}
   \includegraphics[width=0.8\linewidth]{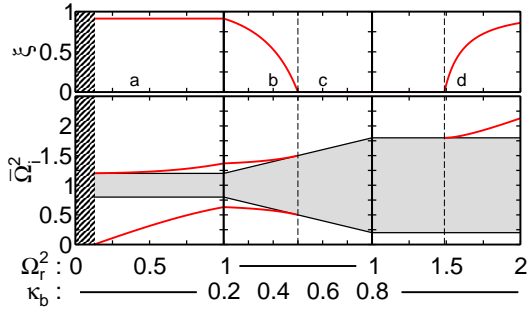}
   \caption{\label{fig:isomod} (Color online) Position $\bar{\Omega}_i$ and total weight $\xi$ of isolated poles of $F(z)$.
  We set $\kappa=1/2$ and change $\Omega_r$, $\kappa_b$ along the dashed path from Fig.~\ref{fig:paramsK2}, i.e.
  from $\Omega_r=0$, $\kappa_b=0.2$ to $\Omega_r=2$, $\kappa_b=0.8$.
  The position of the poles is compared to the continuum of bath modes
  in the interval $\bar{\omega}^2 \in [1-\kappa_b,1+\kappa_b]$,
  filling the grey area around $\bar\Omega^2_i=1$ in the plot. 
  Between the two vertical dashed lines at $\Omega_r=1$, $\kappa_b=1/2$ (left)
and $\kappa_b=0.8$, $\Omega_r=1.4875$ (right)
 no poles exist in agreement with condition~\eqref{NoPoles}. 
  For $\Omega_r^2 \lessapprox 0.12628$, in the dashed region to the left, the positivity  condition~\eqref{PosForChain} is violated and one $\bar{\Omega}_i^2$ becomes negative.
  }
\end{figure}

\begin{figure}[ht]
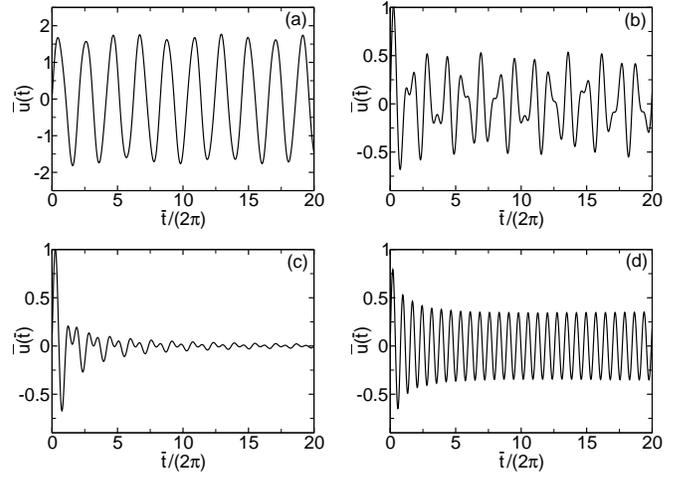

   \includegraphics[width=0.48\linewidth]{Fig6a}
   \hfill
   \includegraphics[width=0.48\linewidth]{Fig6b} \\[1ex]
   \includegraphics[width=0.48\linewidth]{Fig6c}
   \hfill
   \includegraphics[width=0.48\linewidth]{Fig6d}     
   \caption{\label{fig:ut}Function $\bar u(\bar{t})$ for the harmonic chain with $\kappa=1/2$.
   The parameters from panels (a)--(d) correspond to the crosses in Figs.~\ref{fig:paramsK2},~\ref{fig:isomod}.
     They are: 
   (a) $\kappa_b=0.2$, $\Omega_r^2=0.4$ (two poles $\bar{\Omega}_1=0.48$, $\bar{\Omega}_2=1.10$, $\xi_1=0.82$, $\xi_2=0.10$),
   (b) $\kappa_b=0.4$, $\Omega_r^2=1$ (two poles  $\bar{\Omega}_1=0.76$, $\bar{\Omega}_2=1.20$, $\xi_1=0.26$, $\xi_2=0.26$),
   (c) $\kappa_b=0.6$, $\Omega_r^2=1$ (no pole),
   (d) $\kappa_b=0.8$, $\Omega_r^2=1.6$ (one pole  $\bar{\Omega}_1=1.35$, $\xi_1=0.50$).
     }
\end{figure}

As an example let us consider the case $\kappa=1/2$.
The restrictions on the parameters arising from the positivity condition~\eqref{PosForChain} and the stronger condition~\eqref{NoPoles} are summarized in Fig.~\ref{fig:paramsK2}.
We now follow the dashed path in this figure and plot the position $\bar \Omega_{1/2}^2$ of isolated poles and their total weight $\xi=\xi_1+\xi_2$ in Fig.~\ref{fig:isomod}.
Only for parameter combinations in the white unshaded area in Fig.~\ref{fig:paramsK2},
which corresponds to the part between the dashed vertical lines in Fig.~\ref{fig:isomod},
condition~\eqref{NoPoles} is fulfilled.
Accordingly, only panel (c) in Fig.~\ref{fig:ut} (the parameter combination ``c'' in Figs.~\ref{fig:paramsK2},~\ref{fig:isomod})
shows a situation where $\bar{u}(\bar{t}) \to 0$ for $t \to \infty$.
Otherwise, one (parameter combination ``d'') or two (``a'', ``b'') isolated poles exist if one or both  inequalities from~\eqref{NoPoles} are violated. Then, the amplitude of oscillations in $\bar{u}(\bar{t})$ remains finite in the long-time limit.

\begin{figure}[ht]
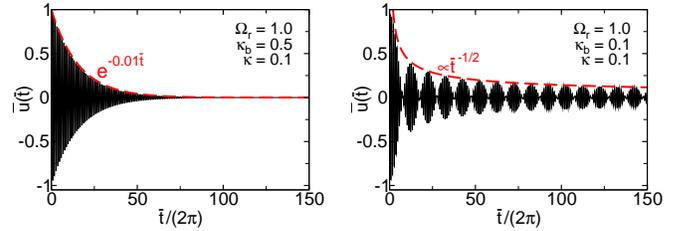

   \includegraphics[width=0.48\linewidth]{Fig7a}
   \hfill
   \includegraphics[width=0.48\linewidth]{Fig7b}
 \caption{\label{fig:ut2} (Color online) Function $\bar u(\bar{t})$ for the inhomogeneous (left panel,
with $\Omega_r=1$, $\kappa_b=0.5$, $\kappa=0.1$)
and homogeneous (right panel,
 with $\Omega_r=1$, $\kappa_b=0.1$, $\kappa=0.1$) harmonic chain at weak damping.
 The dashed red curves indicate the exponential decay from Eq.~\eqref{ChainOhm} (left panel) and the 
 asymptotic decay $\propto 1/\sqrt{\bar{t}}$ of the Bessel function from Eq.~\eqref{HomChainBessel} (right panel).}
\end{figure}

For strong damping situations ($\kappa \sim 1$) shown in Fig.~\ref{fig:ut} 
the function $u(t)$ deviates significantly from an exponentially decaying function, even in the absence of poles (panel (c)). 
Exponential decay occurs
only for weak damping $\kappa \ll \kappa_b$.
For $|\Omega_r^2-1| \ll \kappa_b$ we have 
\begin{equation}\label{ChainOhm}
  \bar{u}(\bar{t})  = \frac{\sin( \Omega_r \bar{t} \,)}{\Omega_r}  \, \exp \Big(\! - \! \frac{\kappa^2}{2 \kappa_b} \bar{t} \Big) \qquad (\kappa \ll 1) \;, 
\end{equation}
as plotted in Fig.~\ref{fig:ut2} (left panel).
Note that the case $\gamma(\Omega) = 0$, 
with an undamped sine function in the weak damping limit, is excluded by the second inequality in Eq.~\eqref{NoPolesWC}.

For the homogeneous chain, with $\Omega_r=1$ and $\kappa_b=\kappa$,
the weak damping limit gives a different result.
Since $\kappa_b = \kappa$, 
the width of the continuum of bath states shrinks to zero for $\kappa \to 0$
such that we do not obtain exponential decay of $u(t)$.
Instead, it is
\begin{equation}\label{HomChainBessel}
 \bar{u}(\bar{t}) =J_0\! \left(\!\frac{\kappa \bar{t}}{2} \right) \,  \sin \bar{t}   \qquad (\kappa \ll 1, \, \mathrm{hom. \; chain})   
\end{equation}
with the Bessel function $J_0(x)$ (cf. Refs.~\cite{Rub60,RH69,Aga71}).
According to condition~\eqref{NoPoles} isolated poles of $F(z)$ cannot occur in this situation.
From the asymptotic behavior of the Bessel function we find that 
here $\bar u(\bar t)$ decays only as $2 (\pi \kappa \bar{t})^{-1/2}$ for $\bar{t} \gg 1$,
as shown in the right panel of Fig.~\ref{fig:ut2}.
Exponential decay in the weak damping limit is only achieved if the coupling $\kappa$ of the central oscillator to the chain becomes small also in comparison to the width ($\sim \kappa_b$) of the continuum of bath states.

\subsection{Thermalization after a quench}

According to the previous discussion,
the central oscillator in the harmonic chain 
equilibrates precisely for parameter combinations that fulfill condition~\eqref{NoPoles}.
We now study, under these conditions, thermalization after a quench that generates a non-thermal environment for the central oscillator (cf. Eq.~\eqref{InitChainT2} below).

\subsubsection{Initial conditions generated by the quench}

We imagine that for $t < 0$ all oscillators are decoupled ($\kappa = \kappa_b=0$) and in thermal equilibrium at temperature $T_0$.
Every oscillator has the same variance
\begin{equation}\label{ChainInitial}
\Omega_b^2 \breve{\Sigma}_{qq}(n) = \breve{\Sigma}_{pp}(n) =  E(T_0, \Omega_b) \;,
\end{equation}
and we do not need to specify further initial expectation values if we are only interested in the stationary state in the long-time limit.

At $t=0$ we quench the system by cranking up the coupling to finite values $\kappa, \kappa_b > 0$. Since $\breve{\Sigma}_{qq}(n)$, $\breve{\Sigma}_{pp}(n)$ do not depend on $n$,
transformation to the normal modes of the bath results in constant functions
\begin{equation}
 \Omega_b^2 \breve{\Sigma}^{(1)}_{QQ}(\omega) = \breve{\Sigma}^{(1)}_{PP}(\omega) =  E(T_0, \Omega_b) 
  \end{equation}
  for the initial bath variances at $t=0$.
The initial bath state is uncorrelated with $\breve{\mathbf{\Sigma}}^{(2)}(\omega_1, \omega_2) = 0$.

According to Sec.~\ref{sec:Equil}, the stationary state in the long-time limit
depends only on the frequency-resolved energy $\breve{\mathcal{E}}(\omega)$ of the initial bath state,
which for the present example is given by the function 
\begin{equation}\label{InitChainT2}
 \breve{\mathcal{E}}(\omega)  = \frac{1+(\omega/\Omega_b)^2}{2} E(T_0,\Omega_b) \;.
\end{equation} 
This function acquires a dependence on $\omega$ through the dispersion of the bath modes after the quench, 
but it does not fulfill Eq.~\eqref{ThermAtWC}.
We thus see that the thermal equilibrium state of uncoupled oscillators before the quench corresponds to a non-thermal state of the coupled chain of oscillators after the quench.
According to condition (T3) from Sec.~\ref{sec:EquilWithList} we expect that the temperature $T_\infty$ of the stationary state depends on the central oscillator frequency $\Omega_r$ even at weak coupling.

From Eqs.~\eqref{CQQLong}--\eqref{CQPLong} or Eqs.~\eqref{CQQLongF},~\eqref{CPPLongF}, the variances in the long-time limit are obtained as
\begin{equation}\label{SigQQChain}
  {\Sigma}_{QQ}^\infty = \frac{1}{2 \Omega_b^2} \left( 1+ \frac{1}{ \Omega_r^2 - \frac{\textstyle \kappa^2}{\textstyle \kappa_b^2} \left(1 - \sqrt{1- \kappa_b^2} \right)} \right) E(T_0,\Omega_b) \;,
\end{equation}
\begin{equation}\label{SigPPChain}
  {\Sigma}_{PP}^\infty =  \frac{1}{2} \left( 1+  \Omega_r^2 \right) E(T_0, \Omega_b) \;.
\end{equation}
We will give further results using normalized quantities
\begin{equation}
 \bar{\Omega}_\infty = \Omega_\infty/\Omega_b \;,
 \quad
 \bar{T}_\infty = T_\infty / \Omega_b \;,
 \quad
 \bar{T}_0 = T_0 / \Omega_b \;. 
\end{equation}
choosing $\Omega_b$ as the unit of energy.

\subsubsection{Thermalization (T2)}

We recall that according to property (T1) the stationary state is always a thermal state of some harmonic oscillator Hamiltonian, such that we should check the stronger property (T2).
From Eq.~\eqref{OmTLong}, the effective frequency associated with the stationary state is
\begin{equation}\label{ChainOm}
  \frac{\bar \Omega_\infty^2}{\Omega_r^2} = \frac{ \Omega_r^2 + 1}{ \Omega_r^2 + \Big[ 1 - \frac{\textstyle \kappa^2}{\textstyle \Omega_r^2 \kappa_b^2} \left(1 - \sqrt{1- \kappa_b^2} \right) \Big]^{-1} } \;.
\end{equation}
We observe that equipartition of energy, i.e. $\bar \Omega_\infty = \Omega_r$, can be achieved only in the weak damping limit $\kappa \to 0$.
For $\kappa > 0$, it is always $\bar \Omega_\infty < \Omega_r$.
This confirms the conditions given for property (T2) in Sec.~\ref{sec:EquilWithList}.

\subsubsection{Thermalization (T3)}

\begin{figure}
 \includegraphics[width=0.8\linewidth]{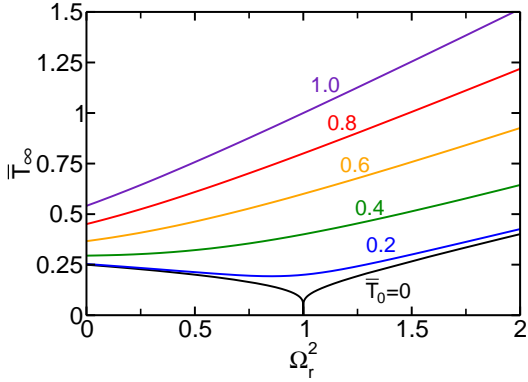}
 \caption{\label{fig:TChain} (Color online) Temperature $\bar{T}_\infty$ of the stationary thermal state at weak damping as given in Eq.~\eqref{TChainWC2}.
 It is shown as a function of $\Omega_r$ for different temperatures $\bar{T}_0 = 0.2, \dots, 0.8  $ of the initial state, as indicated.
 Note that  $\bar{T}_\infty$ does not depend on $\kappa_b$, but the admissible values of $\Omega_r$ for which equilibration occurs are restricted by the second condition in Eq.~\eqref{NoPolesWC} (see also Fig.~\ref{fig:params}). 
 In particular, it must be $0 \le \Omega_r^2 \le 2$.
 }
\end{figure}

For weak damping, Eqs.~\eqref{SigQQChain},~\eqref{SigPPChain} simplify to
\begin{equation}
 \Omega^2  \Sigma_{QQ}^\infty = \Sigma_{PP}^\infty = \frac{1}{2} \left( 1 + \Omega_r^2 \right) E(T_0,\Omega_b) \qquad (\mathrm{for} \; \kappa \to 0) \;.
\end{equation}
Equipartition of energy in the stationary state is evident, and the thermalization (T2) property fulfilled.
To check property (T3), we calculate the temperature 
\begin{equation}\label{TChainWC2}
 \frac{2 \bar T_\infty(\Omega)}{\Omega_r} =  \mathrm{arcoth}^{-1} \Bigg[ \frac{1}{2} \Big( \Omega_r + \frac{1}{\Omega_r} \Big) \coth \Big( \frac{1}{2 \bar T_0} \Big) \Bigg] 
\end{equation}
of the stationary state with Eq.~\eqref{OmTLong} or
the weak damping result~\eqref{TWC}.
We see that $\bar T_\infty(\Omega_r)$ depends explicitly on the central oscillator frequency $\Omega_r$,
as depicted in Fig.~\ref{fig:TChain}.
It is $\bar T_\infty = \bar T_0$ only for $\Omega_r=1$.
As discussed before, 
this results from the fact that $\breve{\mathcal{E}}(\omega)$ after the quench violates condition~\eqref{ThermAtWC}.

We note that $\kappa_b$ does not appear in Eq.~\eqref{TChainWC2}.
In the present example the value of $\kappa_b$ only determines
the admissible values of $\Omega_r$ that lead to equilibration,
as given by the second inequality in Eq.~\eqref{NoPolesWC}.
Once equilibration has been observed,
the temperature of the stationary state at weak damping depends only on the value of $\breve{\mathcal{E}}(\Omega)$ not on
the functional dependence of the spectral function $\gamma(\omega)$.

\subsubsection{The homogeneous chain}

\begin{figure}
\includegraphics[width=0.8\linewidth]{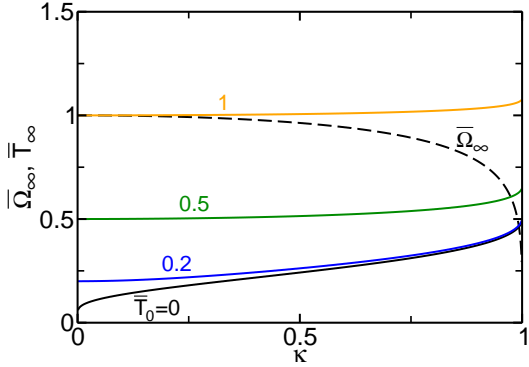}
\caption{\label{fig:HomChain} (Color online) Frequency $\bar{\Omega}_\infty$ (dashed curve) and temperature $\bar T_\infty$ (solid curves) 
for the homogeneous chain, from Eqs.~\eqref{HomChain1},~\eqref{HomChain2}
and shown as a function of $\kappa$.
The temperature curves are plotted for $ \bar T_0 = 0,0.2,0.5,1$ as indicated.}
\end{figure}

For the homogeneous chain with $\kappa_b = \kappa$, $\Omega_r = 1$ Eqs.~\eqref{SigQQChain},~\eqref{SigPPChain} simplify to 
\begin{equation}\label{SigQQHomChain}
  \Sigma_{QQ}^\infty = \frac{1}{2 \Omega_b^2} \left( 1 + \frac{1}{ \sqrt{1 - \kappa^2}} \right)  E(T_0,\Omega_b) \;,
\end{equation}
\begin{equation}\label{SigPPHomChain}
 \Sigma_{PP}^\infty = E(T_0,\Omega_b) \;.
\end{equation}

Equipartition of energy is violated
for any $\kappa > 0$, such that the effective frequency
\begin{equation}\label{HomChain1}
 \bar\Omega_\infty^2 =  \frac{2}{1+(1-\kappa^2)^{-1/2} } 
 \end{equation}
 associated with the stationary state deviates from the central oscillator frequency
 (it is always $\bar{\Omega}_\infty \le 1$).
 The temperature of the stationary state is
 \begin{equation}\label{HomChain2}
 \frac{2 \bar T_\infty}{\bar \Omega_\infty} = \mathrm{arcoth}^{-1} \left[ \coth  \Big( \frac{1}{2 \bar T_0} \Big) \sqrt{\frac{1+ (1-\kappa^2)	^{-1/2}}{2}} \, \right] \;.
\end{equation}
It is $\bar T_\infty > \bar T_0$ for $\kappa > 0$,
for example $\bar T_\infty \to 1/2$ for $\kappa \to 1$ and $\bar T_0 \to 0$ (see Fig.~\ref{fig:HomChain}).

The situation simplifies again in the weak damping limit $\kappa \to 0$,
where we recover from Eqs.~\eqref{SigQQHomChain},~\eqref{SigPPHomChain} the equilibration/thermalization result for the homogeneous chain formulated in Refs.~\cite{RH69,Aga71}:
At weak damping the central oscillator evolves into a stationary thermal state,
with equipartition of energy $\Omega^2  \Sigma_{QQ}^\infty =  \Sigma_{PP}^\infty = E(T_0,\Omega_b)$.
Because of translational invariance this statement applies to every chain oscillator.

We note, however, that thermalization of the homogeneous chain is not perfect.
As discussed in Sec.~\ref{sec:EquilWithList},
observation of a single oscillator in the homogeneous chain
is not sufficient to establish thermalization of the entire chain.
Thermalization fails for a finite chain segment consisting of two or more oscillators, because property (T3) is not fulfilled as seen in Eq.~\eqref{TChainWC2}.
Note that there is no possibility to check property (T3) directly for the homogeneous chain ($\Omega_r=1$ is fixed here), such that results restricted to this situation have to be interpreted carefully~\cite{RH69,Aga71}.

\section{\label{sec:Summary}Conclusion}

Our study of the dissipative quantum harmonic oscillator 
addresses equilibration and thermalization in non-thermal environments.
Equilibration is the generic behavior, which is prevented only in situations where the classical oscillator equation of motion
possesses undamped oscillatory solutions.
The infinite harmonic chain is an example for this behavior.

Thermalization of the central oscillator depends on additional conditions.
Just as for thermal environments,
equipartition of energy requires the weak damping limit but is independent of the precise initial conditions.
The asymptotic temperature $T_\infty$ is obtained from the energy distribution $\breve{\mathcal{E}}(\omega)$ in the initial bath state, and generally depends on the central oscillator frequency $\Omega$.
If we demand that $T_\infty$ is independent of $\Omega$, another condition on $\breve{\mathcal{E}}(\omega)$ follows.
This condition is essential for simultaneous thermalization of several oscillators, 
when a thermal state of the combined system is obtained only if the 
same asymptotic temperature is assumed by each oscillator.

Part of the behavior discussed here generalizes to systems with non-linear interactions.
First, we note that equilibration is possible although the linear system is integrable.
Equilibration occurs because, in a rough sense, the reduced density matrix of the central oscillator involves an average over conserved quantities of the joint oscillator-bath system.
In other words, equilibration of small systems embedded in a large environment 
does not require ergodicity.
Second, because of the linearity and unitarity of quantum mechanical time evolution the stationary state depends explicitly on the initial (bath) state.
But already for the linear system some properties,
such as equipartition of energy, are independent of the initial conditions.
Furthermore, the stationary state depends only
on the energy distribution $\breve{\mathcal{E}}(\omega)$ in the initial bath state.
Effectively, information is lost in the long-time limit and thermalization is possible for a large class of (non-thermal) initial states.

We did neither discuss the generalization of the fluctuation-dissipation relation to the present non-thermal setting, nor the role of stationary non-equilibrium states with finite heat flow that would require coupling to at least two baths with different preparations.
Multi-time correlations functions can be computed within the present formalism, which will allow for the analysis of both issues in the future.

\begin{acknowledgments}

The authors wish to acknowledge helpful discussions with M. Cramer, G.-L. Ingold, and M. Thorwart.
This work was supported by Deutsche Forschungsgemeinschaft through  SFB 652 (B5)
and AL1317/1-2.

\end{acknowledgments}

\appendix

\section{\label{app:Deriv}Operator equations of motion and their solution}

The solution of the dissipative quantum harmonic oscillator model through operator equations of motion instead of transformation to normal modes of $H$
allows for a simple treatment of general initial conditions and time-dependent coefficients.
We here list the relevant steps of the derivation omitted in Sec.~\ref{sec:EOM}, and allow for a time-dependent central oscillator frequency $\Omega(t)$ (cf. Ref.~\cite{ZH95} for a path integral calculation).

The Heisenberg equations of motion $\dot{A}(t) = \rmi [H, A(t)]$ are
\begin{equation} \label{EOMCentral}
 \dot{Q}(t) = P(t) \;, \qquad \dot{P}(t) = -\Omega^2(t) Q(t) - \sum_{\nu=1}^N \lambda_\nu Q_\nu(t) 
 \end{equation}
 for the position and momentum operator of the central oscillator, and 
\begin{equation}  \label{EOMBath} 
 \dot{Q}_\nu(t) = P_\nu(t) \;, \qquad \dot{P}_\nu(t) = - \omega^2_\nu Q_\nu(t) - \lambda_\nu Q(t) 
\end{equation}
for the bath oscillators.

We can read Eq.~\eqref{EOMBath} as an inhomogeneous linear equation for $Q_\nu(t)$. 
Using the Green function for the homogeneous equation $\ddot{Q}_\nu(t)=-\omega_\nu^2 Q_\nu(t)$, we find
\begin{eqnarray}
   Q_\nu(t) &=& \cos \omega_\nu t \, Q_\nu(0) + \frac{1}{\omega_\nu} \sin \omega_\nu t \, P_\nu(0) \nonumber\\
  &&- \lambda_\nu \int_0^t \frac{1}{\omega_\nu} \sin \omega_\nu (t-\tau) Q(\tau) \, \rmd\tau \,.
\end{eqnarray}

Inserting this result into Eq.~\eqref{EOMCentral} gives the equation of motion 
\begin{equation}\label{Qeom}
 \ddot{Q}(t) = - \Omega^2(t) Q(t) + \int_0^t K(t-\tau) Q(\tau) \, \rmd\tau - N(t) 
\end{equation}
for the central oscillator operator $Q(t)$,
with the damping kernel $K(t)$ from Eq.~\eqref{DampKern}
and the noise term
\begin{equation}
 N(t) = \sum\limits_{\nu=1}^N \lambda_\nu \Big( \cos \omega_\nu t \, Q_\nu(0) + \frac{\sin \omega_\nu t}{\omega_\nu} \, P_\nu(0) \Big) \,.
\end{equation}

Eq.~\eqref{Qeom} is an inhomogeneous linear integro-dif\-fer\-en\-tial equation,
which can be solved through solution of the classical equation of motion
\begin{equation}
 \partial_{tt} u(t,t') = - \Omega^2(t) u(t,t') + \int_{t'}^t K(t-\tau) u(\tau,t') \, \rmd\tau \;.
\end{equation}
We need the two solutions $u_1(t,t')$, $u_2(t,t')$ to initial conditions
$u_1(t,t)=1$, $\partial_t u_1(t,t')|_{t=t'}=0$,
and $u_2(t,t)=0$, $\partial_t u_2(t,t')|_{t=t'}=1$.
The solution of the operator equation of motion for $Q(t)$ is then given by
\begin{equation}\label{app:QSol}
 Q(t) = u_1(t,0) Q(0) + u_2(t,0) P(0)  - \int_0^t u_2(t,\tau) N(\tau) \, \rmd\tau \;,
\end{equation}
and it is $P(t) = \dot{Q}(t)$.

With the partial Fourier transforms
\begin{equation}
  \tilde{u}(t,\omega) = \int_0^t u_2(t,\tau) \, \rme^{\rmi \omega \tau} \, \rmd \tau \;,
\end{equation}
\begin{equation}
  \tilde{v}(t,\omega) = \int_0^t \partial_t u_2(t,\tau) \, \rme^{\rmi \omega \tau} \, \rmd \tau \;,
\end{equation}
and the definition of matrices
\begin{equation}\label{UMatTD}
 \mathbf{U}(t) = \begin{pmatrix} u_1(t,0) & u_2(t,0) \\[0.5em] \partial_t u_1(t,0) & \partial_t u_2(t,0)  \end{pmatrix}
\end{equation}
and $ {\mathbf U}(t, \omega) $ as in Eq.~\eqref{UMatOm}
the operators $Q(t)$, $P(t)$ are given by the matrix Eq.~\eqref{QSolMat}.

For constant $\Omega(t) \equiv \Omega$,
the function $u(t)$ used in Sec.~\ref{sec:EOM}
is recovered as $u(t)=u_2(t,0)$, and it is $\dot{u}(t)=u_1(t,0)$
(while $u_1(t,t') \ne \partial_t u_2(t,t')$ for time-dependent $\Omega(t)$).
Then, the partial Fourier transforms $\tilde{u}_2(t,\omega)$ and $\tilde{v}_2(t,\omega)$
are related by Eq.~\eqref{UFourDot} and $\mathbf U(t)$ is given by the simpler
expression~\eqref{UMat}.

Eqs.~\eqref{X},~\eqref{I} and Eqs.~\eqref{Sigma},~\eqref{C} 
for the calculation of expectation values and variances
and Eq.~\eqref{JW} for the construction of the propagating function remain valid for time-dependent $\Omega(t)$.

\section{\label{app:Prop}Propagating function in position representation}

The propagating function in position representation is the Fourier transform
\begin{eqnarray}
   J(q_f,q'_f, q_i,q_i',t) &=& \frac{1}{2\pi} \iint_{-\infty}^{\infty} \rme^{\rmi \tilde{p}(q_f-q_f')} \, \rme^{- \rmi p (q_i -q_i') } \nonumber\\
 &&\times  J_W\Big(\frac{q_f+q_f'}{2},\tilde{p},\frac{q_i+q_i'}{2},p ,t \Big) \, \rmd\tilde{p} \, \rmd p \; \nonumber\\
\end{eqnarray}
of Eq.~\eqref{JW}.
It results in the expression
\begin{eqnarray}\label{JPos}
   J(Y,y,&&X,x,t) = \frac{|j_6|}{2 \pi} \exp \Big[  j_1 x^2 + j_2 xy + j_3 y^2 \nonumber\\
 &&+ \rmi \Big( (j_4 x  + j_5 y) X + (j_6 x  +j_7 y) Y + j_8 x + j_9 y   \Big)   \Big] \;, \nonumber\\
\end{eqnarray}
where we write $Y=(q_f+q_f')/2$, $y=q_f-q_f'$,
$X=(q_i+q_i')/2$, $x=q_i-q_i'$ for abbreviation and drop the time argument in $j_k \equiv j_k(t)$.
The $9$ real parameters $j_1, \dots, j_9$ in this expression are related to the parameters of $J_W(\tilde{\mathbf{x}},\mathbf{x},t)$ in Eq.~\eqref{JW} through 
\begin{eqnarray}\label{JPosParams}
 j_1 &=& - \dfrac{C_{QQ}}{2 U_{QP}^2} \;, \quad
 j_2 = - \dfrac{C_{QP}}{U_{QP}} +  C_{QQ} \dfrac{U_{PP}}{U_{QP}^2} \;, \nonumber\\
 j_3 &=& - \dfrac{1}{2} C_{PP} - \dfrac{U_{PP}^2}{2 U_{QP}^2} C_{QQ} + \dfrac{U_{PP}}{U_{QP}} C_{QP} \nonumber\\
 j_4 &=& \dfrac{U_{QQ}}{U_{QP}} \;, \quad
 j_5 = U_{PQ} - \dfrac{U_{QQ} U_{PP}}{U_{QP}} \;, \quad
 j_6 = - \dfrac{1}{U_{QP}} \;, \nonumber\\
 j_7 &=& \dfrac{U_{PP}}{U_{QP}} \;, \quad
 j_8 = \dfrac{1}{U_{QP}} I_Q \;, \quad
 j_9 = I_P - \dfrac{U_{PP}}{U_{QP}} I_Q \;. \nonumber\\
\end{eqnarray}
Explicit insertion of $\mathbf{U}(t)$ from Eq.~\eqref{UMat} or~\eqref{UMatTD} gives expressions 
that allow for direct comparison with the literature.
For example, the expressions given in Ref.~\cite{ZH95} are recovered for $\mathbf{I}(t) \equiv 0$ such that the terms $j_8 x$, $j_9 y$ vanish.

Obviously,
the position representation leads to less convenient expressions for the propagating function,
and obscures the clear formal structure of Eq.~\eqref{JW}.
In particular, the expressions~\eqref{JPosParams} are singular for $u(t) \to 0$, which gives a complicated representation of the $\delta$-distribution for the propagating function at $t=0$ and $t \to \infty$ instead of the simple limit for $J_W(\tilde{\mathbf{x}},\mathbf{x},t)$ (cf. Eq.~\eqref{JWLong}).

\end{document}